\documentclass[aps,prb,english,twocolumn,showpacs,preprintnumbers,amsmath,amssymb,floatfix,superscriptaddress,longbibliography]{revtex4-1}

\usepackage[T1]{fontenc}
\usepackage[latin9]{inputenc}  
\usepackage{graphicx}
\usepackage{amssymb}
\usepackage{babel}

\usepackage{hyperref}
\hypersetup{
     colorlinks = true,
     citecolor  = blue,  
     linkcolor  = blue 
}

\makeatletter
\usepackage[version=3]{mhchem}
\usepackage{color}

\global\arraycolsep=2pt
\usepackage{stmaryrd}
\usepackage{amsmath}
\usepackage{amssymb}
\usepackage{graphicx}
\usepackage{textcomp}
\usepackage{calrsfs}
\usepackage{yfonts}
\usepackage{bm}
\usepackage{color}
\newcommand{\ben}{\begin{equation*}}
\newcommand{\een}{\end{equation*}}
\newcommand{\bean}{\begin{eqnarray*}}
\newcommand{\eean}{\end{eqnarray*}}

\newcommand{\be}{\begin{equation}}
\newcommand{\ee}{\end{equation}}
\newcommand{\bea}{\begin{eqnarray}}
\newcommand{\eea}{\end{eqnarray}}

\makeatother
\usepackage{babel}
\makeatother 
\usepackage{color}

\begin{document}

\title{Understanding ice and water film formation on soil particles by 
combining DFT and Casimir-Lifshitz forces}

\author{M.  Bostr{\"o}m}
\email{mathias.bostrom@ensemble3.eu}
\affiliation{Centre of Excellence ENSEMBLE3 Sp. z o. o., Wolczynska Str. 133, 01-919, Warsaw, Poland}

\author{S. Kuthe}
\email{kuthe@kth.se}
\affiliation{Department of Materials Science and Engineering, KTH Royal Institute of Technology, SE-100 44 Stockholm, Sweden}

 \author{S. Carretero-Palacios}
 \affiliation{Departamento de F\'isica de Materiales and Instituto de Materiales Nicol\'as Cabrera, Universidad Aut\'onoma de Madrid, 28049 Madrid, Spain}
\affiliation{Present Address: Instituto de Ciencia de Materiales de Madrid, ICMM-CSIC, C/ Sor Juana In\'es de la Cruz, 3,  28049 Madrid, Spain}

 \author{V. Esteso}
\affiliation{Departamento de F\'isica de la Materia Condensada, ICMSE-CSIC, Universidad de Sevilla, Apdo. 1065, Sevilla, Spain}
\affiliation{European Laboratory for Non-Linear Spectroscopy (LENS), Via Nello Carrara 1, Sesto F.no 50019, Italy} 

\author{Y. Li}
  \affiliation{Department of Physics, Nanchang University, Nanchang 330031, China}
  \affiliation{Institute of Space Science and Technology, Nanchang University, Nanchang 330031, China}

\author{I. Brevik}
\affiliation{Department of Energy and Process Engineering, Norwegian University of Science and Technology, NO-7491 Trondheim, Norway}

\author{H. R. Gopidi}
\affiliation{Centre of Excellence ENSEMBLE3 Sp. z o. o., Wolczynska Str. 133, 01-919, Warsaw, Poland}
  
\author{O. I. Malyi}
\affiliation{Centre of Excellence ENSEMBLE3 Sp. z o. o., Wolczynska Str. 133, 01-919, Warsaw, Poland}

\author{B. Glaser}
\email{bjoerng@kth.se}
\affiliation{Department of Materials Science and Engineering, KTH Royal Institute of Technology, SE-100 44 Stockholm, Sweden}

\author{C. Persson}
\email{claspe@kth.se}
\affiliation{Department of Materials Science and Engineering, KTH Royal Institute of Technology, SE-100 44 Stockholm, Sweden}
\affiliation{Centre for Materials Science and Nanotechnology, Department of Physics, University of Oslo, P. O. Box 1048 Blindern, NO-0316 Oslo, Norway}

\begin{abstract}
{Thin films of ice and water on soil particles play crucial roles in environmental and technological processes. Understanding the fundamental physical mechanisms underlying their formation is essential for advancing scientific knowledge and engineering practices. Herein, we focus on the role of the Casimir-Lifshitz force, also referred to as dispersion force, in the formation and behavior of thin films of ice and water on 
soil particles at 273.16\,K, arising from quantum fluctuations of the electromagnetic field and depending on the dielectric properties of interacting materials. We employ the first-principles density functional theory (DFT) to compute the dielectric functions for two model materials, CaCO$_3$ and Al$_2$O$_3$, essential constituents in various soils. These dielectric functions are used with the Kramers--Kronig relationship and different extrapolations to calculate the frequency-dependent quantities required for determining forces and free energies.  Moreover, we assess the accuracy of the optical data based on the DFT to model dispersion forces effectively, such as those between soil particles.
Our findings reveal that moisture can accumulate into almost micron-sized water layers on the surface of calcite (soil) particles, significantly impacting the average dielectric properties of soil particles. This research highlights the relevance of DFT-based data for understanding thin film formation in soil particles and offers valuable insights for environmental and engineering applications.}
\end{abstract}
\date{\today}

\maketitle

\section{Introduction}
\label{AllIntroduction}
Ice and water, omnipresent in nature, play pivotal roles in an array of environmental and technological phenomena, as evidenced by multiple studies.\cite{JacobGlaciers2012,LannuzelNatClimCha2020,WEBER199759} Therefore, comprehending the primary physical principles that dictate the formation of thin ice and water films is crucial for numerous scientific pursuits and engineering applications. One specific example lies within civil engineering, where thin films of ice and water on soil particles bear significant implications.\cite{HAMADA2023} They influence the construction and maintenance of critical infrastructure, including building foundations, roads, and bridges. Gaining insights into the generation and characteristics of these films enables engineers to design structures with enhanced resistance to frost heave and thaw settlement damage. Likewise, soil is a complex and dynamic system, and understanding its behavior at a fundamental level can lead to new insights and discoveries in geology, chemistry, and physics. Comprehension about the formation of thin films of ice and water on soil particles is also critical for predicting and mitigating the impact of climate change on soil ecosystems. As temperatures fluctuate, the formation and melting of ice and water films affect the availability of nutrients and water to plants, the stability of soil structure, and overall, the health of soil micro-organisms.\,\cite{TiessenNature1994} In addition, having knowledge of the formation of thin films of ice and water on soil particles is important for advancing our understanding of basic questions such as frost heave,\cite{Rempel2007_frostheave} and, more generally, the physical and chemical properties of soil. In recent years, there has been growing interest in the role of the Casimir-Lifshitz force in the formation and behavior of thin films of ice and water in diverse (astro-)geological systems, covering   ice seeding particles in clouds\,\cite{LUENGOMARQUEZMacDowell2021,luengo2022WaterIce} and the potential involvement of insulating gas hydrate caps\,\cite{BostromEstesoFiedlerBrevikBuhmannPerssonCarreteroParsonsCorkery2021} in facilitating the persistence of liquid water on celestial bodies like the moon Enceladus.\,\cite{Waite2006} This force, which arises from quantum fluctuations of the electromagnetic (EM) field, also called dispersion force, strongly depends on the dielectric properties of the interacting materials, amongst other parameters. 

\par
Motivated by the above, herein, we investigate the necessity of accurate dielectric functions derived from density functional theory (DFT), which can provide more information of the optical response of materials than the standard experimental measurements, for reliable modeling of dispersion forces between soil particles. Specifically, we present the imaginary part of the dielectric functions (related to dissipative properties of the materials) for CaCO$_3$ and Al$_2$O$_3$, vital components found in diverse soil compositions.\cite{Muhs2001,Nordt2010} These are then used with a Kramers-Kronig relationship and different extrapolations to calculate the real-valued dielectric function evaluated at imaginary frequencies,which facilitates the computation. This latter quantity is used to calculate forces and free energies. Our main objective is to determine how well-established low and medium-energy optical spectra from DFT can be combined with high-energy extrapolations, aiming to confirm the validity of previous conclusions based on the comparison between experimental optical data and theoretical forces.\cite{ParsegianNinham1969} Remarkably, our findings indicate that the calculated interaction energies remain largely unaffected by the specific approach employed for the low and high-energy extrapolations in a few significant scenarios.

Our DFT-based predictions indicate a dielectric constant of 8.7 for calcite, which aligns closely with the previously measured static dielectric constant range of 8 to 9.\,\cite{LebronRobinsonGoldbergLeschCalcite2004} However, our current research reveals a significant phenomenon: the accumulation of moisture, in the form of water molecules, in micron-sized layers on the surface of calcite particles found in soil. This accumulation profoundly impacts the average dielectric properties of soil particles. Notably, existing models\,\cite{LebronRobinsonGoldbergLeschCalcite2004} utilized to estimate water content in soils rely on a mineral static dielectric constant value of 5 as an input parameter. This poses a potential problem since calcite is a primary constituent in various soils. Consequently, accurate modeling of soil dielectric properties necessitates a comprehensive understanding of these properties for constituent materials such as calcite, quartz, water, and others.\,\cite{LebronRobinsonGoldbergLeschCalcite2004} Given the significant impact of calcite on various soil compositions, addressing this issue becomes imperative.

\section{The semi-classical theory for Lifshitz interactions}
\label{MathiasReflections}

\subsection{Some initial considerations}
\label{MathiasInitialConsiderations}
 \par The semi-classical theory of intermolecular forces follows from the realization that much of the quantum electrodynamics formalism\,\cite{Lifshitz1956,Dzya} can be derived via Maxwell's equations with boundary conditions, and the subsequent assignment to each quantized EM mode a zero point energy at zero temperature (or, at finite temperatures, the free energy). Previously, it was believed that the complex Lifshitz theory\cite{Dzya} required knowledge of the dielectric function over the entire spectrum to calculate dispersion forces in layered structures. However, van Kampen, Nijboer, and Schram,\cite{KampenNijboerSchram1968} made progress in simplifying the theory by demonstrating the derivation of non-retarded interactions from a semi-classical approach. In 1969, Parsegian and Ninham,\cite{ParsegianNinham1969} further advanced this work, leading to numerous Lifshitz and Casimir interaction calculations. Although outdated in light of subsequent publications, Parsegian and Ninham's pioneering paper,\cite{ParsegianNinham1969} remains significant. Their breakthrough was recognizing that only partial knowledge of the optical spectrum of different materials is sufficient to understand the van der Waals-Lifshitz interaction between planar surfaces separated by intervening material.\cite{ParsegianNinham1969,NinhamParsegianWeiss1970,Richmond_1971_magnetic,BarashGinzburg1975} In what follows, we will explore similar concepts to examine how approximations in DFT-based material properties present in soil particles relate to the accuracy of calculated Hamaker constants and Lifshitz interactions. Our findings confirm that different high-frequency extrapolations for evaluating DFT-derived dielectric functions are not crucial for obtaining accurate Lifshitz forces and Hamaker constants.

 \subsection{Optical quantities and their interrelationships}
 \label{Optics}

The real ($\varepsilon_i'$) and imaginary ($\varepsilon_i''$) parts of the dielectric function (for material $i$ = 1, 2, and 3) are related via the well-known Kramers-Kronig relationships using Cauchy principal $(P)$ value integration\,\cite{landau2013statistical}
\begin{equation}
\varepsilon_i'(\omega)=1+\frac{2}{\pi} {{P}}\int_0^\infty d\Omega \frac{ \Omega \, \varepsilon_i''(\Omega)}{\Omega^2-\omega^2}. 
\end{equation}
 We can also use the well-known relationship to the refractive index $n_i$ and the extinction coefficient $k_i$\,\cite{landau2013statistical} 
 \begin{equation}
\sqrt{\varepsilon_i'(\omega)+i \varepsilon_i''(\omega)}=n_i(\omega)+i k_i(\omega),
\end{equation}
which can be rewritten as
\begin{equation}
\varepsilon_i'(\omega)=n_i(\omega)^2-k_i(\omega)^2,
\end{equation}
and
\begin{equation}
\varepsilon_i''(\omega)=2n_i(\omega)k_i(\omega).
\end{equation}
The Kramers-Kronig transformation requires a sufficiently wide frequency range to obtain accurate estimates for the complex-valued dielectric function. {\it Ab initio} DFT modeling of the  has the advantage over utilizing experimental data by describing the response functions for much larger frequencies. Normally they are calculated from 0 to $\sim 1.5\times 10^{17}$\,rad/s (i.e., $\sim100$\,eV),
while the measurements are restricted to photon energies typically below some tens of eV and by using different sources for the exciting beam. Moreover, the wider frequency range is important also to accurately calculate the Casimir-Lifshitz forces.

\subsection{The Ninham-Parsegian model for Lifshitz forces}
\label{Mathias_on_NinParsModel}

 \par We are revisiting the Ninham-Parsegian\cite{ParsegianNinham1969} model for Lifshitz interaction, closely following the approach outlined in some remarkably lucid papers from the 1970s.\,\cite{ParsegianNinham1969,NinhamParsegianWeiss1970,Richmond_1971_magnetic} We aim to enhance our understanding of this model and its implications. 
 At zero temperature, the non-retarded ($NR$) van der Waals-Casimir-Lifshitz interaction energy is simply the change in zero-point energies  ($\hbar \omega_\lambda/2$) of the allowed quantized EM surface modes when two surfaces are at a finite distance $d$ compared to when the surfaces are infinite apart,
\begin{equation}
E^{NR}(d)=\frac{\hbar}{2} \sum_\lambda \int \frac{d^2 q}{(2 \pi)^2}  [\omega_\lambda(d)-\omega_\lambda(\infty)], \label{Eq_Ch1_sec:5:5b}
\end{equation}
where zero-point energies are summed over different allowed ($\lambda$) modes and integration is done over the different wavevectors ($q$).
The EM surface modes arise from solving Maxwell's equations with appropriate boundary conditions. The dispersion equation to be solved to obtain the surface modes is in the non-retarded regime (i.e. short separations between the plates, where one can ignore finite velocity of light),\cite{NinhamParsegianWeiss1970}
\begin{equation}
    D(d,\omega)=1-\frac{[\varepsilon_1(\omega)-\varepsilon_2(\omega)] [\varepsilon_3(\omega)-\varepsilon_2(\omega)]}{[\varepsilon_1(\omega)+\varepsilon_2(\omega)] [\varepsilon_3(\omega)+\varepsilon_2(\omega)]} e^{-2 q d}=0
    \label{Eq_Ch1_sec:5:5}
\end{equation}
with subscripts 1 and 3 representing the two interacting materials, through material 2. 

\par Following the generalized argument theorem\,\cite{NinhamParsegianWeiss1970} (see the equations below), a much simplified formula for intermolecular forces between surfaces can be presented. Assuming an analytic function $\Delta$ with zeros at $\omega_\lambda (d)$, and that has a derivative which has singularities at $\omega_\lambda (\infty)$, complex analysis produces,\cite{NinhamParsegianWeiss1970}
\begin{equation}
\sum_\lambda \frac{\hbar}{2} [\omega_\lambda(d)-\omega_\lambda(\infty)]=\frac{1}{2 \pi i} \oint_C \frac{\hbar \omega}{2}  \frac{d \omega}{\Delta(d,\omega)} \frac{\partial \Delta(d,\omega)}{\partial \omega}.
    \label{Eq_Ch1_sec:5:6}
\end{equation}
Here, $C$ is the closed path going down the imaginary axis and closing in the right-hand plane forming a semi-circle at which all quantities vanish. Doing a partial integration, we find,\cite{NinhamParsegianWeiss1970}
\begin{equation}
\frac{1}{2 \pi i} \oint_C \frac{\hbar \omega}{2}  \frac{d \omega}{\Delta(d,\omega)} \frac{\partial \Delta(d,\omega)}{\partial \omega}=\frac{\hbar}{4 \pi} \int_{-\infty}^\infty d\xi \ln [\Delta(d,i \xi)],
    \label{Eq_Ch1_sec:5:7}
\end{equation}
where $\xi$ is a variable of integration  (which at finite temperatures it goes over to the so-called Matsubara frequencies).
By substituting the above expression in Eq.(\ref{Eq_Ch1_sec:5:5b}) we get,
\begin{equation}
E^{NR}(d)\approx\frac{\hbar}{4 \pi^2} \int_0^\infty dq q\int_0^\infty d\xi \ln [ 1-\Delta_{12}^{NR} \Delta_{32}^{NR}  e^{-2 q d}],
    \label{Eq_Ch1_sec:5:8}
\end{equation}
where the non-retarded (NR) reflection coefficients are given as
\begin{equation}
\Delta_{ij}^{NR}=\frac{\varepsilon_i(i \xi)-\varepsilon_j(i \xi)}{\varepsilon_i(i \xi)+\varepsilon_j(i \xi)}.
    \label{Eq_Ch1_sec:5:9}
\end{equation}
This can be generalized when the finite speed of light is accounted for by writing it as a sum of a transverse magnetic (TM) and a transverse electric (TE) contributions,\cite{NinhamParsegianWeiss1970} 
\begin{equation}
E(d)\approx\frac{\hbar}{4 \pi^2} \int_0^\infty dq q\int_0^\infty d\xi \{\ln [ G^{TM}]+\ln [ G^{TE}]\},
    \label{Eq_Ch1_sec:5:10_ret}
\end{equation}

\begin{equation}
G^{TM/TE}= 1-\Delta_{12}^{TM/TE} \Delta_{32}^{TM/TE} e^{-2 \gamma_2 d},
    \label{Eq_Ch1_sec:5:10_TM}
\end{equation}

\begin{equation}
\Delta_{ij}^{TM}=\frac{\gamma_j \varepsilon_i(i \xi)-\gamma_i \varepsilon_j(i \xi)}{\gamma_j \varepsilon_i(i \xi)+\gamma_i \varepsilon_j(i \xi)},\,\,\Delta_{ij}^{TE}=\frac{\gamma_j -\gamma_i }{\gamma_j +\gamma_i },
    \label{Eq_Ch1_sec:5:9_TMTE}
\end{equation}
where $\gamma_i^2=q^2+\xi^2 \varepsilon_i/c^2$, and $c$ is the speed of light. 

\par At finite temperature, $T$, the zero point energy of each mode should be replaced with the Helmholtz free energy,\cite{NinhamParsegianWeiss1970}
\begin{equation}
F(\omega,T)=k_B T\ln[2 \sinh(\hbar \omega/[2 k_B T])].
    \label{Eq_Ch1_sec:5:11}
\end{equation}
 with $k_B$ the Boltzmann constant. 
 A derivative of the Helmholtz free energy expression, arising from a partial integration in the same way as in Eq.(\,\ref{Eq_Ch1_sec:5:7}), provides a factor $\coth[\hbar \omega/(2 k_B T))]$. The $\coth$ factor has an infinite number of poles on the imaginary axis. This leads to the fact that zero temperature and finite temperatures can be dealt with a simple substitution,\cite{NinhamParsegianWeiss1970}
\begin{equation}
\frac{\hbar}{2 \pi} \int_{0}^\infty d \xi \rightarrow k_BT\sum_{m=0}^\infty{}^{\prime}, 
  \xi\rightarrow \xi_m=2 \pi k_B T m/\hbar,
    \label{Eq_Ch1_sec:5:13}
\end{equation}
where the sum was originally from minus infinity to plus infinity leading to a factor of 1/2 for the $m$=0 term.
The quantity related to forces expressed in Matsubara frequencies, $\xi_m$, can be obtained directly from $\varepsilon_i'(\omega)$ and $\varepsilon_i''(\omega)$, i.e. from materials optical properties, via well-known Kramers-Kronig relationships\,\cite{landau2013statistical}
\begin{equation}
\varepsilon_i(i \xi_m)=1+\frac{2}{\pi}\int_0^\infty d\omega \frac{ \omega \varepsilon_i''(\omega)}{\omega^2+\xi_m^2}, 
\label{KramKronEq}
\end{equation}
This quantity is real-valued and decays smoothly towards one leading to very simple calculations.

\par The leading non-retarded  interaction energy (using Eq.(\,\ref{Eq_Ch1_sec:5:8})) is,\,\cite{ParsegianNinham1969}
\begin{equation}
E^{NR}(d)\approx\frac{-A}{12 \pi d^2},
    \label{Eq_Ch1_sec:5:2:1}
\end{equation}
with $A$ a Hamaker constant for the system. We will use the finite temperature non-retarded expression for the Hamaker constant,
\begin{equation}
A = -6 {k_B T} \sum_{m=0}^\infty{}^{\prime} \int_0^\infty dq q \ln[ 1-\Delta_{12}^{NR} \Delta_{32}^{NR} e^{-2 q}].
    \label{Hamaker}
\end{equation}

\par  The measurements by Haydon and Taylor\,\cite{HaydonTaylor1968} of the energy were in the past\,\cite{ParsegianNinham1969}  used to estimate the Hamaker constant for water surfaces separated by a biomolecular lipid film. The measured energy was $-3.94\times10^{-6}$\,J/m$^2$ for a film of estimated thickness 56\,\AA. To test the Lifshitz theory, one requires to model dielectric functions that are derived from optical data or, in recent years, density functional theory (DFT) has become a commonly employed method.  Within the Ninham and Parsegian model\,\cite{ParsegianNinham1969} the dielectric functions $\varepsilon(\omega)$ of water, oil (resembling the lipid membrane), and many different materials can be  modeled as,
\begin{equation}
\varepsilon(\omega)=1+\frac{c_{rot}}{1-i \omega/\omega_{rot}}+\sum_j \frac{c_{j}}{1- (\omega/\omega_j)^2+i \gamma_j \omega},
        \label{Eq_Ch1_sec:5:2:2}
\end{equation}
where $\omega_j$ are characteristic frequencies and $c_j$ are proportional to the oscillator strengths. For calculations of the Hamaker constant, one  requires the dielectric functions for imaginary frequencies,
\begin{equation}
\varepsilon(i \xi)=1+\frac{c_{rot}}{1+ \xi/\omega_{rot}}+\sum_j \frac{c_{j}}{1+ (\xi/\omega_j)^2},
        \label{Eq_Ch1_sec:5:2:3}
\end{equation}
where the damping term ($\gamma_j$) on the imaginary frequency axis can usually be ignored since bandwidths are generally much smaller 
than the absorption frequencies. The rotational relaxation ($\omega_{rot}$) occurs at very low frequencies. In the far ultraviolet, both water and the oil film behave as a simple plasma with
\begin{equation}
\varepsilon(i \xi)=1+\frac{\omega_P^2}{\xi^2},
        \label{Eq_Ch1_sec:5:2:4}
\end{equation}
being $\omega_P^2=4 \pi N e^2/m$ the plasma frequency with $N$, $e$, and $m$ the electron density, charge, and mass, respectively. Since water and oil have similar electron densities,\,\cite{ParsegianNinham1969} contributions from the ultraviolet frequency region and higher were ignored. The experimental Hamaker constant for water surfaces separated by a biomolecular lipid film was (within large error estimates) equal to $A\sim4.66\times10^{-21}$\,J.\cite{HaydonTaylor1968,ParsegianNinham1969}
Testing different parameters\,\cite{ParsegianNinham1969} suggested that removing the ultraviolet contribution gave a Hamaker constant  $A\sim3.9\times10^{-21}$\,J, while varying the refractive index of oils lead to $A\sim4.5-5.4\times10^{-21}$\,J.

 Since the 1970s, numerous groups worldwide have conducted extensive comparisons between theory and experiments analyzing the effect of the accuracy of the optical data in the calculation of dispersion forces. Despite this, the fundamental concepts remain largely unchanged. In the following discussion, we will demonstrate that for a few selected model examples, the treatment of the extrapolated high-frequency tail in DFT-based dielectric functions for imaginary frequencies is not as critical as initially anticipated when it comes to accurately describing Hamaker constants and Lifshitz interactions.

\section{DFT modeling of the solids}
\label{SudhansuBjornClas}
The modeling of Al$_2$O$_3$ and CaCO$_3$, essential components in diverse soil compositions, are performed within the DFT and with the projector augmented wave (PAW) method for the $GW$-type core potentials, as implemented in the Vienna Ab initio Simulation Package (VASP).\cite{VASP1999} 
The valence configurations for the atoms are chosen to 
C: $2s^2p^2$, 
O: $2s^2p^4$,
Al: $2s^2p^63s^2p^1$, and
Ca: $3s^2p^64s^2$.  
As these compounds are wide-gap insulators, we employ as default the generalized gradient approximation, revised exchange-correlation functional for solids (PBEsol), developed by Perdew, et al.;\cite{PBEsol2008} the band gap energy is corrected with a hybrid functional. 
The unit cells are described by ten-atom trigonal lattices, and  
the irreducible Brillouin zones are sampled by a $6\times 6 \times 6 $ {\bf k}-mesh. 
A quasi-Newton (variable metric) algorithm is utilized for the structural relaxation
with a cut-off energy of 800\,eV, to an accuracy of $10^{-4}$\,eV/\AA\ for the forces on all atoms, 
Thereafter, the charge density is generated with a 600\,eV cut-off energy, 
using the linear tetrahedron integration, and iterated in the electronic self-consistent 
loop to reach an energy accuracy of $10^{-6}$\,eV.
The irreducible representations of the electronic eigenstates are determined by the 
open-source program  Irvsp.\cite{Gao2021}

From the electronic structure, the imaginary part $\varepsilon''(\omega)$ of the macroscopic dielectric function is calculated. With the independent single-electron eigenfunctions, the response due to electronic transitions is described as the joint density-of-states modulated by the optical matrix elements.
In the long-wavelength limit, the latter reads
\begin{align}
 \lefteqn{\varepsilon''^{\,ele}_{\alpha\alpha}(\omega) =\lim_{{\bf q} \rightarrow 0}  
\frac{4\pi^2e^2}{V_{\Omega} q^2}
 \sum_{v,c,{\bf k}}  \delta (\epsilon_{c,{\bf k}} - \epsilon_{v,{\bf k}} - \hbar \omega) }\nonumber\\
 &\qquad \qquad \qquad  \times\langle   u_{c,{\bf k}+{\bf e}_{\alpha}q} | u_{v, {\bf k} }   \rangle     
  \langle   u_{v, {\bf k} } |  u_{c,{\bf k}+{\bf e}_{\alpha}q}   \rangle \,,
\label{eq:eps2a}
\end{align}
in the three Cartesian directions ${\bf e}_{\alpha}$. Here, $V_{\Omega}$ is the unit-cell volume and $u_{v/c}$ is the cell periodic part of the valence ($v$) or conduction ($c$) state eigenfunction with the energy $\epsilon_{v/c,{\bf k}}$. Local field effects are neglected. As the two compounds are insulators, we perform the {\bf k}-space summation by Bl\"{o}chl's linear tetrahedron method. Since the accuracy of the calculation can strongly depend on the size of the  {\bf k}-point grid,\cite{CROVETTO2016} we use a $12\times 12 \times 12 $ {\bf k}-mesh though the values of the low-frequency dielectric constants are  sufficiently converged already for the charge density from the $6\times 6 \times 6 $ {\bf k}-mesh.   

Alumina and calcite are ionic compounds, and we, therefore, consider the local lattice dynamics. The vibrations associated with the longitudinal optical (LO) modes  build up an electric field that screens the carriers. The dipole-active LO phonons and the corresponding transverse optical (TO) modes contribute to the dielectric response. In the  long-wavelength limit, the phonon dispersion is approximated to be constant, and the ionic response is modeled as Lorentz oscillators. 
\begin{equation}
\varepsilon''^{\,ion}_{\alpha\alpha}(\omega) =  
\sum_j \frac{ S_j\,\omega_{\rm TO}^2 \Gamma_j \omega}
{(\omega_{\rm TO}^2 - \omega^2)^2 + \Gamma_j\omega^2}\, .
\label{eq:eps2b}
\end{equation}
$\Gamma_j$ is the damping and $S_j$ is the oscillator strength of the $j$th mode in its vibration direction. We employ the density functional perturbation theory to compute the Hessian matrix of the ionic displacements, incorporating the symmetry of the crystals. 

The total imaginary part of the dielectric function is the summation of the two contributions. 
The corresponding dielectric response function for the real part of $\varepsilon'(\omega)$ 
is obtained from the Kramers--Kronig relation. For the average response functions, we take the
arithmetic mean of the three Cartesian directions.

\section{Results}
\label{results}
\subsection{Crystalline structures and dielectric response functions of alumina and calcite }
Both Al$_2$O$_3$ and CaCO$_3$ crystalize in the space group structure R$\overline{3}$c  
($D_{3d}^6$; No. 167), based on the ditrigonal-scalenohedral point group with 
rhombohedral Bravais lattices. Since the accuracy of split-off energies in the electronic 
a structure can depend on bond lengths and bond angles,\cite{Persson1999} we relax 
the crystalline structures with four different exchange-correlation functionals; see Table.~\ref{tab:DFTrel},
where the two lattice constants describe hexagonal lattices.
\begin{table}[ht!]
\caption{Lattice constants $a$ and $c$ of alumina and calcite describing the hexagonal structures. 
The direct band-gap energy $E_{g,\Gamma}^{dir}$ refers to the $\Gamma$-point.}
 \renewcommand{\arraystretch}{1.2}
\begin{tabular}{ l  rrrrr} 
 \hline  \hline 
    &  LDA  & PBE  & PBEsol & HSE & Expt. \\
 \hline 
Al$_2$O$_3$ &&&& &\\
	\hspace{0.3 cm} $a $ [\AA ] & 4.728 & 4.809 & 4.777 & 4.742 &       4.7657$\,^{\cite{dAmour1978}}$\\ 
&&&& &4.761$\,^{\cite{Lucht2003}}$\\ 
&&&& &4.7597$\,^{\cite{Oetzel1999}}$\\ 
	\hspace{0.3 cm} $c $ [\AA ] & 12.884 & 13.122 & 13.018 & 12.950 &    13.010$\,^{\cite{dAmour1978}}$\\
 &&&& &13.011$\,^{\cite{Lucht2003}}$\\
 &&&& &12.993$\,^{\cite{Oetzel1999}}$\\
	\hspace{0.3 cm} $E_{g,\Gamma}^{dir}$ [eV] & 6.45 & 5.85 & 6.02 & 8.7 & 8.8$\,^{\cite{French1990}}$\\
  &&&& &9.1$\,^{\cite{Bortz1989}}$\\
CaCO$_3$ &&&& &\\
	\hspace{0.3 cm} $a $ [\AA ]   & 4.939 & 5.039 & 4.990 & 4.990 & 4.988$\,^{\cite{Maslen1995}}$\\
 &&&& &4.9899$\,^{\cite{Graf1961}}$\\
  &&&& &4.984$\,^{\cite{Karunadasa2019}}$\\
	\hspace{0.3 cm} $c $ [\AA ] & 16.302 & 17.225 & 16.841 & 17.097 & 17.068$\,^{\cite{Maslen1995}}$\\
 &&&& &17.064$\,^{\cite{Graf1961}}$\\
  &&&& &17.056$\,^{\cite{Karunadasa2019}}$\\
	\hspace{0.3 cm} $E_{g,\Gamma}^{dir}$ [eV] & 5.67 & 5.63 & 5.70 & 8.0 &  \,\,\,\,\,\,\,\,5.65-6.35$\,^{\cite{Baer1993}}$\\
 &&&&   &   6.9-7.7$\,^{\cite{Vos2015}}$\\\\
 \hline  \hline  
\end{tabular}                     
\label{tab:DFTrel}
\end{table}

As expected, the local density approximation (LDA) overbinds by about 1\%, while the
regular generalized gradient approximation (PBE) underbinds by about 1\%. Both the revised 
PBE for solids (PBEsol) and the hybrid functional (HSE with 30\% Hartree-Fock exchange)
agree very well with the experimental data. Since we want to compute the electronic 
transitions on a dense {\bf k}-mesh and to energetically very high states, we choose 
to use the PBEsol functional. 
Lattice parameters for the rhombohedral lattices are $a = 5.137 $\,\AA\ and $\gamma = 55.34$\,$^\circ$ 
for Al$_2$O$_3$ and $a = 6.307 $\,\AA\ and $\gamma = 46.57$\,$^\circ$ for CaCO$_3$, with 
the PBEsol potential.
Although the two oxides crystallize in the same space group symmetry, they have rather 
different crystalline structures (Supplemental Material (SM)\cite{SM2023}), which mainly depends 
on the cation sizes and valence configurations. For alumina, each O atom binds to two 
Al atoms with the bond length 1.86\,\AA\ and to two other 
Al atoms with the bond length 1.97\,\AA .
For calcite, each O atom has a bond to one 
C  atom  with the bond length 1.29\,\AA\ and to two 
Ca atoms with the bond length 2.34\,\AA . 

The underestimated gap energy for the PBE potential is adjusted by 
a constant energy shift of the conduction bands so that the $\Gamma$-point gap 
corresponds to that of HSE. The two compounds are wide-gap insulators, and we do not
expect any valence-conduction band hybridization that could otherwise affect the band 
dispersion,\cite{Persson2006} and thereby also the transition probability.
The differences in the bond character are reflected in the electronic structures.
Al$_2$O$_3$ is a insulator with a direct gap at the $\Gamma$-point, and we estimate the
gap energy to $E_g^{dir} \approx  $ 8.7\,eV. CaCO$_3$, on the other hand, has an indirect
gap of $E_g^{ind} \approx  $ 7.3\,eV, located close to the {\bf k}-point (\textonehalf,\,\textonehalf,\,0),
for which the direct gap is $E_g^{dir} \approx  $ 7.4\,eV.

Further, both compounds have the same single-group irreducible representations at the $\Gamma$-point for the 
energetically lowest conduction state (a single degenerate $\Gamma_1^+$) and the topmost valence state 
(single degenerate $\Gamma_2^-$). 
However, while the second highest valence state in alumina is a double degenerate state ($\Gamma_3^-$) 
only $0.04$\,eV below the topmost valence state, the corresponding state in calcite is single degenerate 
($\Gamma_2^+$) 0.48\,eV below the topmost $\Gamma$-point valence state. 
The irreducible representations for three symmetry points are presented in SM\cite{SM2023}.
Calcite has more flat conduction band dispersion, and one could expect a stronger onset to the 
electronic dielectric response for this compound.
\begin{figure}[h!]
  \centering
  \includegraphics[width=1.0\columnwidth]{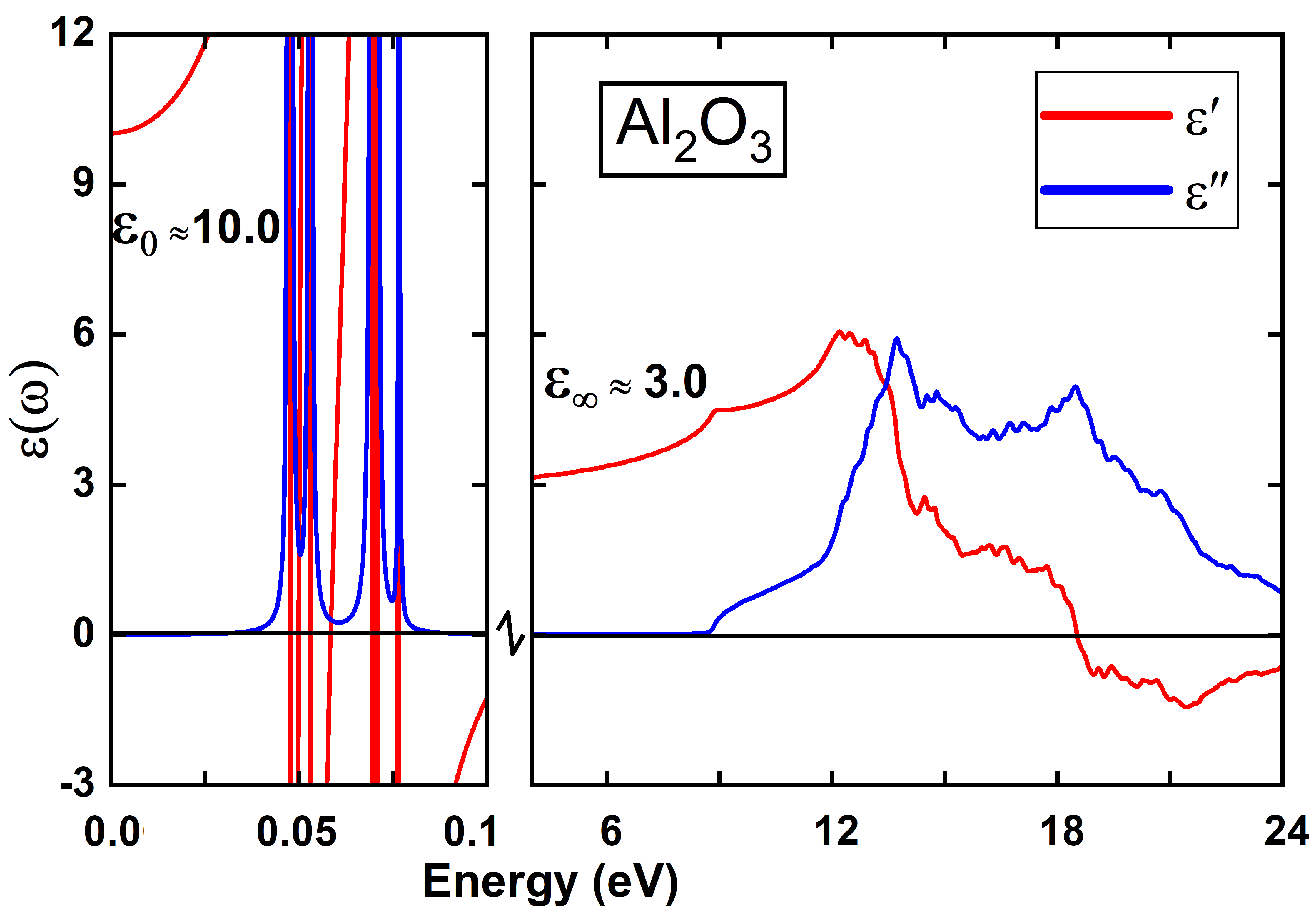}
  \includegraphics[width=1.0\columnwidth]{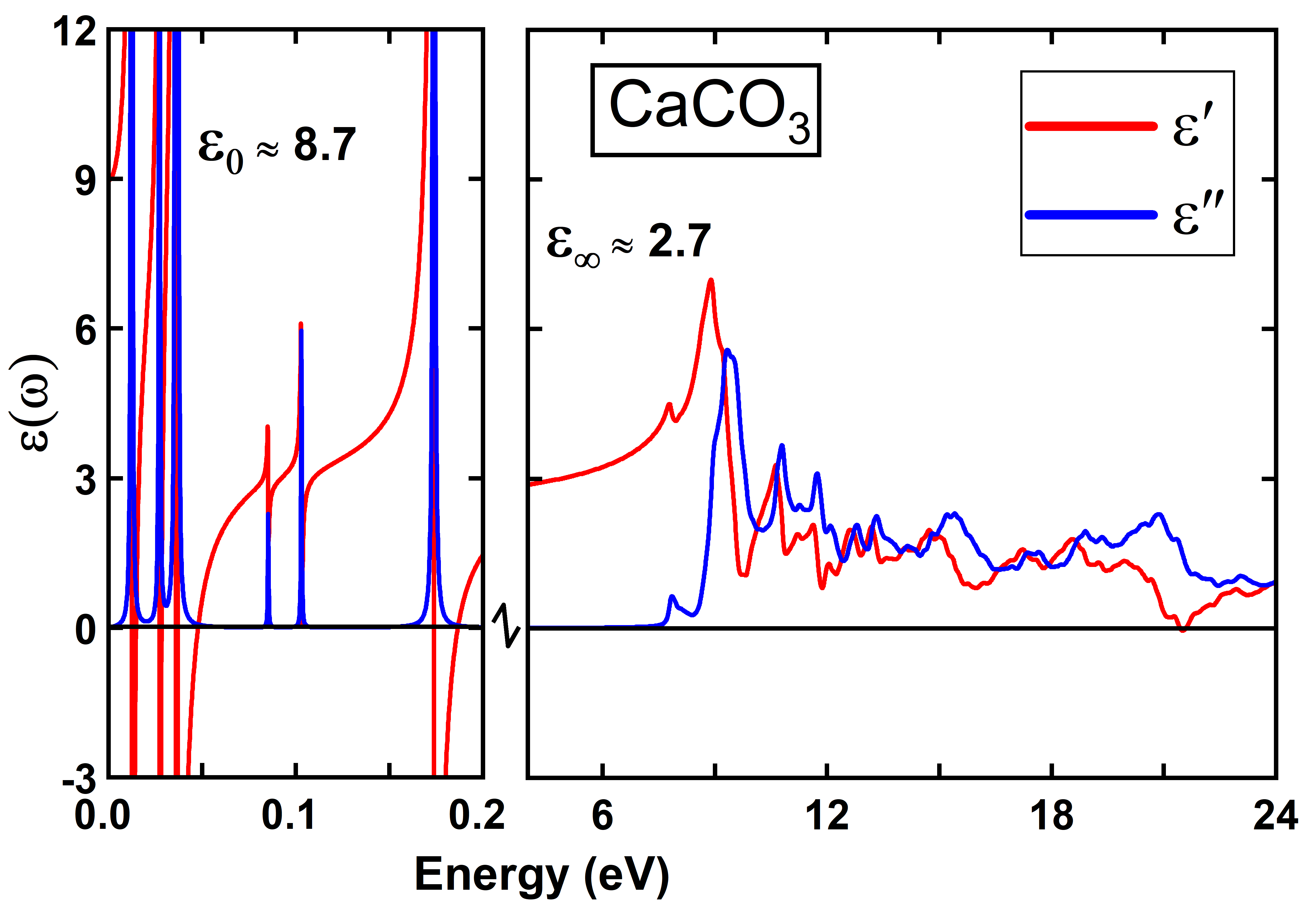}
  \caption{ (Color online) 
  Dielectric function of Al$_2$O$_3$ and CaCO$_3$. Left panel describes the vibrational contribution
  and the right panel describes the electronic transitions.}
  \label{fig:dft_diel}
\end{figure}

The dielectric response functions for alumina and calcite are presented in Fig.~\ref{fig:dft_diel}. 
The response due to electronic transitions contributes above the direct-gap energy on the eV scale,
while the lattice dynamics contributes on the 0.1-eV scale; here, below 0.2\,eV.
We find that the PAW potential for Al with the electronic valence configuration $2s^2p^63s^2p^1$ easily 
yields incorrect vibrational frequencies, and we, therefore, instead use the corresponding potential with 
the  valence configuration $3s^2p^1$. As expected, CaCO$_3$ has a strong electronic response right 
above 9\,eV, while Al$_2$O$_3$ has a smoother increase of the response up to the energy 14\,eV.
Since CaCO$_3$ constitutes both a lighter and a heavier cation, it is natural that the compound has
vibrations that are both lower and higher in frequency than those of Al$_2$O$_3$. 
From the figure, one can observe that the two compounds have rather different dielectric functions, both in 
the regime of the vibrational contribution and for higher frequencies where the electronic transitions contribute. 
In SM,\cite{SM2023}, we present the dielectric functions in more detail for both CaCO$_3$ and Al$_2$O$_3$, 
showing that there is only a moderate difference between the components in the perpendicular and 
the parallel directions. That is also obvious for the dielectric constants (Table~\ref{tab:eps}).
Both compounds have high-frequency constants from the electronic contribution that is close to 3. 
The optical vibration modes contribute to the static dielectric constant; this is 
somewhat larger for Al$_2$O$_3$ ($\sim7.0$) compared to CaCO$_3$ ($\sim6.0$). 
One can notice that the main difference is that Al$_2$O$_3$ has a much larger response in the
parallel direction. 
Overall, there is a good agreement with the experimental findings,~\cite{Young1973} and  
the average static dielectric constant is calculated to be 10.0 in Al$_2$O$_3$, and 8.7 in CaCO$_3$.
\begin{table}[h]
\caption{The static $\varepsilon_0$ and high-frequency $\varepsilon_\infty$  dielectric constants
in the perpendicular, $\perp$, and parallel, $||$, ($z$-) direction. The average ($avg.$) data 
represents the arithmetic mean.}
 \renewcommand{\arraystretch}{1.2}
\begin{tabular}{ l  rrrrr} 
 \hline  \hline 
    & & DFT       &   & Expt.~\cite{Young1973}  &  \\
    &  $\perp$  &   $||$ &\hspace{0.3cm} $avg.$  & \hspace{2.0cm} $\perp$  &  $||$\\
 \hline 
Al$_2$O$_3$ &&&&& \\
\hspace{0.3 cm} $\varepsilon_0$  & 9.3&11.4&10.0  &9.34 &11.54\\
\hspace{0.3 cm} $\varepsilon_\infty$  & 3.0&  3.0&  3.0  && \\
CaCO$_3$ &&&&&\\
\hspace{0.3 cm} $\varepsilon_0$  & 9.2& 7.8& 8.7   & 8.68, 8.5 
&\hspace{0.4cm}8.31, 8.0\\
\hspace{0.3 cm} $\varepsilon_\infty$ & 2.9& 2.3& 2.7 && \\
 \hline  \hline  
\end{tabular} 
\label{tab:eps}
\end{table}

\subsection{Sensitivity of DFT-based Hamaker constants for selected calcite systems} 

Before we set out to exploit the calculated dielectric functions at imaginary frequencies, we first explore how sensitive the results are to the low and high-frequency extrapolations methods. 

The dielectric response for the Matsubara frequencies $\varepsilon(i \xi_m)$ is obtained from 
the Kramers--Kronig relation as in Eq.\,\ref{KramKronEq}.    
The calculated electronic structure from DFT implies energies up to about 250 eV, much higher than energies reached by standard experimental measurements with ellipsometers (typically, well below 10 eV). Higher energies have only small probability for excitation. However, in order to include high-energy transitions, we extrapolate the high-frequency tail of $\varepsilon(i \xi_m)$ with a $1/\xi_m^2$ behaviour up to about 10 keV. With that extrapolation we can analyze the accuracy of the calculations of the Hamaker constants with respect to the number of Matsubara frequencies. 
Moreover, while the default calculations include semicore states down to 100 eV below the valence band maximum, we have also generated spectra of the dielectric functions with no semicore states for the cations Al and Ca, i.e., $3s^2p^1$ and $3p^64s^2$ valence configurations, respectively. Those spectra are denoted by "no SC".  

To analyze the importance to consider the optical phonons for small Matsubara frequencies (typically, $m < 4$), we have generated diectric functions for which the vibrational contribution is completely neglected 
(those spectra are denoted by "no vib") but where the static dielectric constant is included for the $m = 0$ term (denoted by "no vib with $\varepsilon_0$"). 

Parsegian and Ninham's study\,\cite{ParsegianNinham1969} in 1969 provided clear evidence that having partial knowledge of the optical spectra could sometimes be enough to make reasonable estimates of Hamaker constants and calculate the corresponding force. Assuming accurate model calculations using DFT, the extrapolation schemes mentioned earlier in this discussion yield very similar Hamaker constants for the calcite-ice-vapor system, with an accuracy of approximately 10\%. According to the findings presented in Table\,\ref{tb1}, the predicted values for the calcite (1)-ice (2)-vapor (3) configuration are approximately A$_{123}\sim-3.39\times 10^{-20}$\,J and A$_{123;0}\sim0.26\times 10^{-20}$\,J,
where A$_{123;0}$ is the contribution from the zeroth Matsubara term of A$_{123}$.

Significant variations in precision are observed when calculating the Hamaker interaction for calcite-vapor-calcite at different temperatures, as indicated in Table\,\ref{tb2}. At lower temperatures, a greater number of Matsubara terms is required to cover the necessary upper-frequency range for achieving comparable accuracy. Utilizing only 500 Matsubara terms could result in a substantial 25\% error in the calculated Hamaker constant. In Table\,\ref{tb2}, the "default", "noSC", and "no vib with $\varepsilon_0$" approximations all yield identical zero frequency Hamaker constants (as shown in the first row of Table\,\ref{tb2}). However, when vibrations are disregarded, the zero frequency Hamaker constant experiences a decrease of 0.045, 0.185, and 0.749$\times 10^{-20}$\,J at temperatures of 70\,K, 370\,K, and 1500\,K, respectively. It is crucial to highlight, however, that when considering different material combinations, such as cases where the dielectric functions of the materials intersect at certain frequencies or for the large separation behavior of gapped metals,\,\cite{MathiasRizwanHarshanIverClasSashaPRB2023} the results become more reliant on the chosen approximations.

\begin{table}[h!]
\centering
\begin{tabular}{l|r|c|cc}
  \hline
    \hline
  model & $m_{max}$ &  $A_{123}$ ($\rm 10^{-20} J$) & $A_{123;0}$ ($\rm 10^{-20} J$)   \\
  \hline
   default & 250                         & -3.088 & 0.260 \\
   &500                                  & -3.332 & 0.260 \\
    &1000                                & -3.385 & 0.260 \\
      &1500                              & -3.391 & 0.260 \\
        &2000                            & -3.392 & 0.260 \\
        &&& \\ 
    no SC &2000                          & -3.459 & 0.260 \\
    no vib &2000                         & -3.325 & 0.306 \\
    no vib with $\varepsilon_0$ &2000    & -3.371 & 0.260 \\ 
  \hline
\end{tabular}
\caption{\label{tb1} The Hamaker constants at 273.16\,K using Eq.\ref{Hamaker}, $A_{123}$ and its contributions from the zeroth Matsubara term $A_{123;0}$ for various three-layer configurations with CaCO$_3$ (1)-ice (2)-vapor (3) for different models CaCO$_3$ and using different cut-off Matsubara number ($m_{max}$). }
\end{table}

\begin{table}[h!]
\centering
\begin{tabular}{c|c|c|c|c}
  \hline
   \hline
  model &$m_{max}$ &  70\,K&  370\,K&  1500\,K  \\
   \hline
  && $A_{123}$ ($\rm 10^{-20} J$) & $A_{123}$ ($\rm 10^{-20} J$)& $A_{123}$ ($\rm 10^{-20} J$) \\
  \hline
   default & 0  &0.0505 &        0.267                   &1.082 \\
  & 500                      & 10.441 &   14.316    &14.970 \\
    &1000                            & 13.051 &   14.430    &14.972 \\
    &1500                            & 13.802 &   14.442    &14.972 \\
    &2000                            & 14.078 &   14.445    &14.972 \\
  \hline
   \hline
\end{tabular}
\caption{\label{tb2} The Hamaker constants at different temperatures using Eq.\,\ref{Hamaker}, $A_{123}$ ($\rm 10^{-20} J$), in three-layer configurations with CaCO$_3$ (1)-vacuum (2)-CaCO$_3$ (3) using different cut-off Matsubara number ($m_{max}$) for the default dielectric function. The case with $m_{max}$=0 corresponds to the zero frequency Hamaker constant.}
\end{table}

\subsection{Parameterised $\varepsilon(i\xi)$ for optimised data sets for calcite and alumina}
\label{HarshanSasha}

To enable simple use of the calculated dielectric functions to study, for example, Casimir-Lifshitz interactions, we  present  parametrized average dielectric functions  (see Table\,\ref{tbHarshanSasha} for parameters) using a 14-mode oscillator model,\cite{Sasha2016} exploiting Eq.\,\ref{Eq_Ch1_sec:5:2:3} but without any rotational relaxation. To be explicit we use the following model.

\begin{equation}
\varepsilon(i \xi)=1+\sum_j \frac{C_{j}}{1+ (\xi/\omega_j)^2}.
        \label{ParameteriseddielEq}
\end{equation}
Here $\omega_j$ are the characteristic frequencies (given in $eV$ in the Table\,\ref{tbHarshanSasha}) and $C_j$ are proportional to the oscillator strengths. For ice and cold water ($T = 273.16$\,K)  we use parameterised dielectric functions given in the literature.\cite{JohannesWater2019,LUENGOMARQUEZMacDowell2021,luengo2022WaterIce}

\begin{table*}
\centering
\caption{Parametrization of the average dielectric function of continuous media, $\varepsilon(i \xi)$, at
imaginary frequencies for Al$_2$O$_3$ and CaCO$_3$ as calculated with first-principles calculations. In this table frequencies are given in eV. The largest difference between fitted and calculated $\varepsilon(i\xi)$ is about 0.08$\%$. \label{tbHarshanSasha} }

\begin{tabular}{ p{2.3cm} |p{2.3cm} |p{2.3cm} |p{2.3cm}}
\hline
\hline
\multicolumn{4}{c}{ $C_j$ and $\omega_j$ (in eV) for different compounds}  \\
\hline
\multicolumn{2}{c}{Al$_2$O$_3$}  & \multicolumn{2}{c}{CaCO$_3$}\\
\hline
modes ($\omega_j$) & coefficient ($C_j$) & modes ($\omega_j$) & coefficient ($C_j$)\\
\hline
0.0478 & 3.4263 & 0.0038 & 0.0879 \\ 
0.0684 & 3.5999 & 0.0127 & 2.9661 \\ 
1.1552 & 0.0015 & 0.035 & 2.6176 \\ 
13.0704 & 1.0213 & 0.1696 & 0.4434 \\ 
20.5561 & 0.8539 & 1.6088 & 0.0023 \\ 
48.8508 & 0.0929 & 10.3375 & 0.8039 \\ 
119.7988 & 0.0295 & 18.9299 & 0.6241 \\ 
1288.9534 & 0.0 & 34.3621 & 0.2317 \\ 
67441.835 & 0.0005 & 71.2998 & 0.0 \\ 
102566.9502 & 0.0 & 81.1893 & 0.0221 \\ 
407868.9726 & 0.0 & 84.7809 & 0.0 \\ 
889915.6853 & 0.0 & 114.126 & 0.0 \\ 
1723890.6517 & 0.0 & 124.3659 & 0.0 \\ 
3447781.2791 & 0.0 & 241.5762 & 0.0005 \\ 
\hline
\hline
\end{tabular}
\end{table*}

 \subsection{Casimir-Lifshitz force near alumina and calcite surfaces}
\label{MathiasCalciteAluminaComparison}

\begin{figure}
  \centering
  \includegraphics[width=1.0\columnwidth]{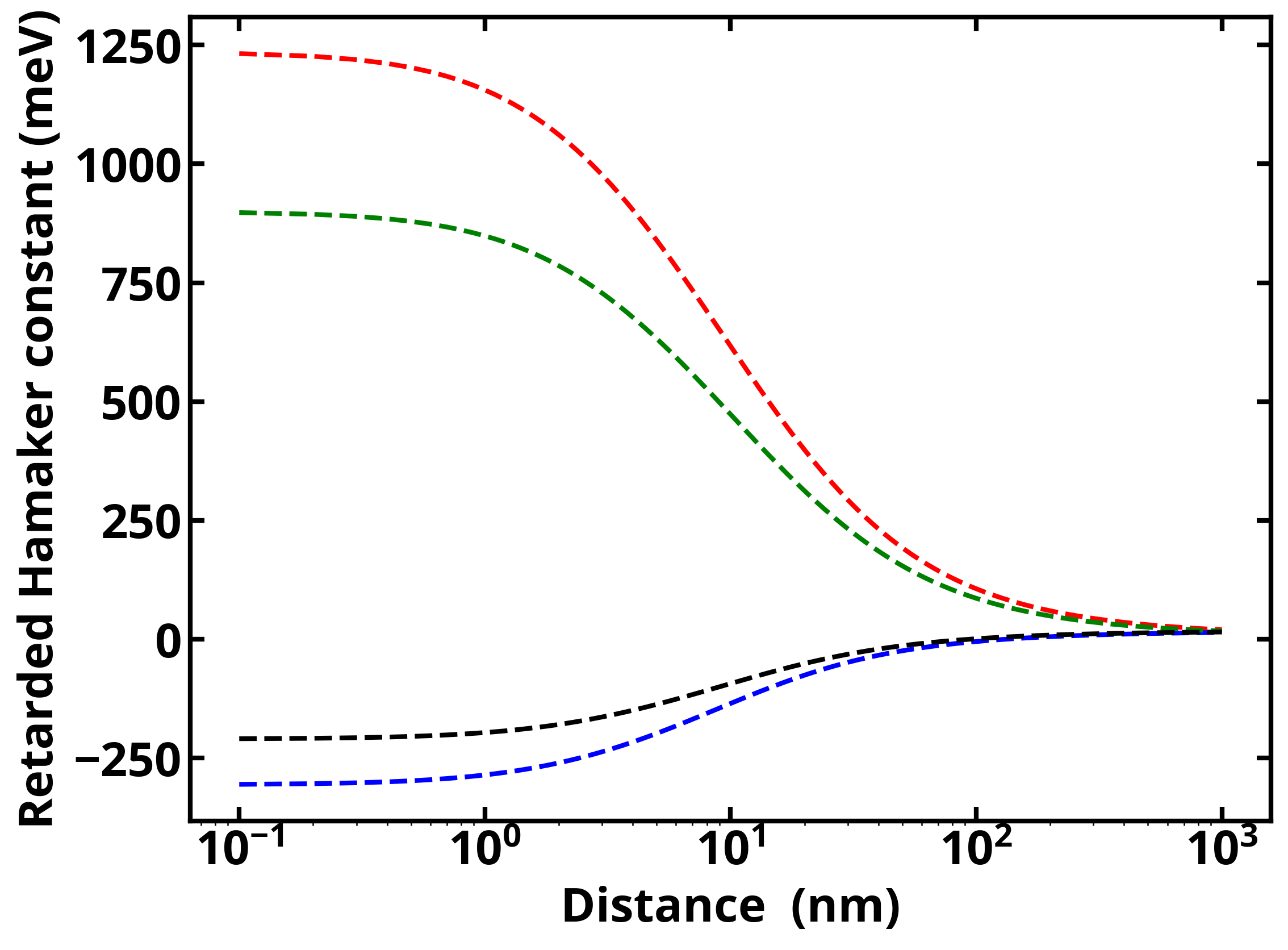}
  \caption{\label{RetardedHamaker} (Color online) The retarded Hamaker constant, $A^{ret}(d)=-F(d,T)\times12 \pi d^2$, for alumina-vacuum-alumina (red curve), calcite-vacuum-calcite  (green curve), alumina-water-vapor  (blue curve), and calcite-water-vapor (black curve). Temperature is 273.16\,K, and other details are given in the text. The corresponding free energies in the region where blue and black curves cross over to positive values are studied in Fig.\,\ref{RetardedFreeEnergy}.}
\end{figure}

\begin{figure}
  \centering
    \includegraphics[width=1.0\columnwidth]{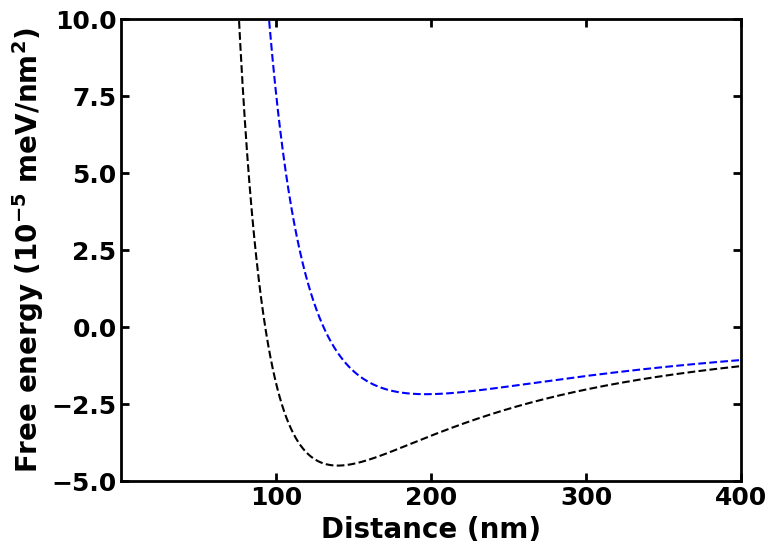}
  \caption{\label{RetardedFreeEnergy} (Color online) The retarded free energy per unit area for two cases studied in Fig.\,\ref{RetardedHamaker} where energy minima are predicted: alumina-water-vapor  (blue curve), and calcite-water-vapor (black curve). Temperature is 273.16\,K, and other details are given in the text.
 }
\end{figure}

The relationship between the retarded (distance-dependent) Hamaker constant and the retarded free energy, denoted as $A^{ret}(d)$ and $F(d,T)$ respectively, can be expressed as $A^{ret}(d)=-12\pi d^2\times F(d,T)$. This connection is illustrated in Figure \ref{RetardedHamaker} for various material combinations, including alumina-vacuum-alumina, calcite-vacuum-calcite, alumina-water-vapor, and calcite-water-vapor. The first two cases unequivocally confirm the well-known phenomenon where the interaction between identical surfaces is attractive and influenced by the material properties of the surfaces. Moreover, in situations involving the interface between a solid and a region with water vapor, there is a possibility of a short-range repulsion transitioning into a long-range attraction, that enables the formation of thin water films. In the latter two cases, where water serves as an intermediate layer, theoretical analysis suggests that in the presence of moisture (water vapour), a thin layer of water can indeed form on the outer surface of calcite, such as on soil particles. The corresponding free energy for these cases are presented in Figure \ref{RetardedFreeEnergy}. Interestingly, the systems exhibit energy minima for finite-sized water layers, further supporting the notion of water formation at these interfaces.
At short separations, specifically when the finite velocity of light can be considered infinite, the product of reflection coefficients is proportional to
\begin{equation}
    \frac{(\varepsilon_1-\varepsilon_2) (\varepsilon_3-\varepsilon_2)}{(\varepsilon_1+\varepsilon_2) (\varepsilon_3+\varepsilon_2)}.
    \label{ratio_reflection}
\end{equation}

This observation suggests that in cases where the dominating frequency range exhibits dielectric functions which fulfill $\varepsilon_1>\varepsilon_2>\varepsilon_3$, a repulsive interaction can occur. On the other hand, when the intermediate layer possesses a higher (or lower) dielectric function than both surrounding media within the dominant frequency range, an attractive force emerges. Analyzing the dielectric functions of water (which has an exceptionally high zero frequency dielectric constant) and calcite (whose dielectric function surpasses that of water at intermediate and high frequencies), we find that, through energy minimization, a Casimir-Lifshitz force can induce the formation of moisture on the surfaces of calcite (and alumina). This phenomenon, driven by Casimir-Lifshitz interactions, leads to the wetting of soil particles by water. In the subsequent subsection, we will further discuss how this Casimir-Lifshitz-induced water wetting affects the effective dielectric function of calcite soil particles.

\subsection{Application to water and ice formation on calcite surfaces}
\label{MathiasCalcite}

\begin{figure}
  \centering
  \includegraphics[width=1.0\columnwidth]{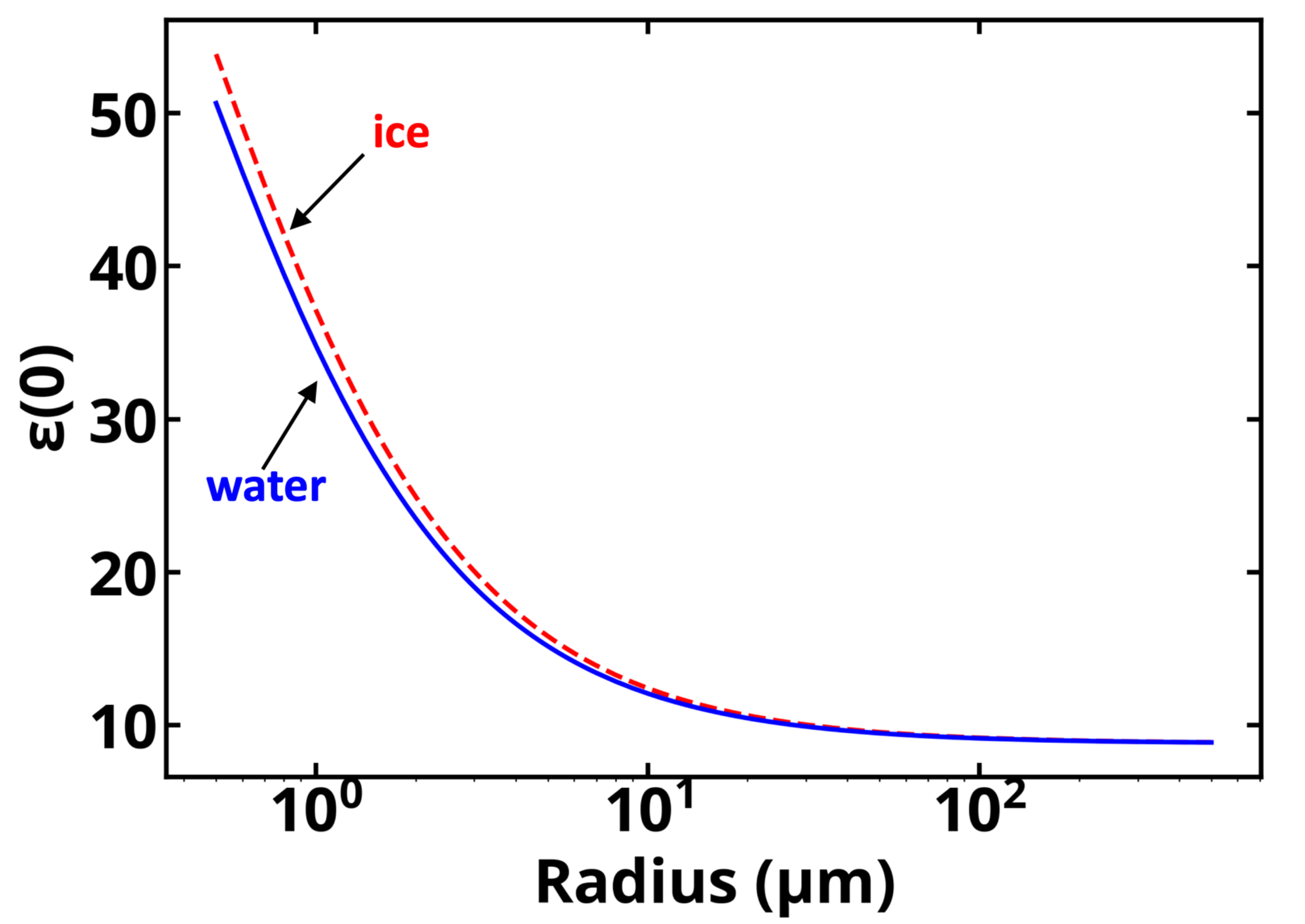}
  \caption{\label{EffectiveDielectricConstants} (Color online) The estimated effective dielectric constants for ice-coated calcite sphere (red dashed curve) and for water-coated calcite sphere (blue solid curve), both curves as functions of the bare calcite radius. Details are given in the text.}
\end{figure}

In a pioneering paper, Elbaum and Schick\,\cite{Elbaum} predicted that a minimum in the dispersion free energy of a thin liquid water film growing on ice would provide an explanation for observed partial melting on ice surfaces near the triple point of water. Following improvements of the modeling of water and ice dielectric functions,\cite{JohannesWater2019,LUENGOMARQUEZMacDowell2021,luengo2022WaterIce} further works explored this idea. These studies include ice melting and ice formation on ice nucleating particles in the atmosphere,\cite{LUENGOMARQUEZMacDowell2021,luengo2022WaterIce} the modeling for anomalous stability of gas hydrates in ice cold water,\cite{BostromEstesoFiedlerBrevikBuhmannPerssonCarreteroParsonsCorkery2021,LiCorkeryCarreteroBerlandEstesoFiedlerMiltonBrevikBostrom2023} and ice formation/melting\,\cite{JohannesWater2019,Esteso4layerPCCP2020} on cold water surfaces. 
Some additional effects of temperature and intermolecular forces on ice adhesion have been discussed by Emelyanenko {\it{et al.}}.\cite{EmelyanenkoEmelyanenkoBoinovich2022}
An interesting idea to explore is the understanding of how the accumulation of ice-cold water or ice from available water vapor outside a calcite surface occurs at the triple point of water. This phenomenon is predicted to lead to the formation of either a thin water or ice film, resulting in a reduction of the overall free energy. Additionally, we identify a previously overlooked correction to the effective dielectric function of soil particles associated with this process. Addressing this aspect is crucial for the advancement of soil science models. To investigate this phenomenon, we utilize the dielectric functions for ice and cold water proposed by Luengo-M\'arquez and  MacDowell.\cite{LUENGOMARQUEZMacDowell2021} The prediction based on Casimir-Lifshitz theory is that the growth of almost micron-sized ice or water layers is favored. For each calcite model considered in Table\,\ref{tb1}, the combination of calcite-ice-vapor results in an equilibrium ice layer with thickness d$_2^{eq}$. Utilizing the retarded finite temperature Casimir-Lifshitz theory, the values for d$_2^{eq}$ are found to be 0.151 $\mu$m, 0.153 $\mu$m, 0.121 $\mu$m, and 0.141 $\mu$m for the "default" calculations, "default noSC", "no vib", and "no vib with $\varepsilon_0$", respectively. These estimated thicknesses of approximately 0.12 $\mu$m to 0.15 $\mu$m are not significantly influenced by the number of terms included in the Matsubara summation, as long as a minimum of 500 terms is considered. 
Using the best available models for the dielectric functions of calcite and cold water,\cite{LUENGOMARQUEZMacDowell2021} we find that above the freezing point of water, the Lifshitz interaction promotes the existence of water vapor accumulating on the calcite surface leading to an $\sim0.14\,\mu$m water film on the surface of soil particles. As an example, a calcite particle with a radius of $\sim1\,\mu$m (neglecting curvature effects for demonstration purposes) could experience a volume increase of around 48\% due to the presence of the wetting film. This phenomenon has implications for the effective dielectric function of soil, even in the absence of liquid water but in contact with water vapor. Notably, based on volume-averaged theory, the effective dielectric constant of a water-coated calcite particle would be approximately 34.8 for the specific example provided. Similarly, an ice coating with a thickness of approximately 0.15 $\mu$m would result in an effective dielectric constant of approximately 37.1. Figure \ref{EffectiveDielectricConstants} illustrates the estimated effective dielectric constants for calcite spheres coated with either ice or water as a function of the calcite particle's radius. Above the freezing point of water, water can adsorb onto soil particles, while at the triple point of water, the growth of ice, water, or a combination thereof depends on the initial conditions. It is noteworthy that even for a calcite particle with a radius of 100 $\mu$m, the effective dielectric constant is enhanced by approximately 4\% compared to the dielectric constant of pure calcite.

\section{Conclusions}
\label{conclusions}

Lebedew, in 1894, likely pioneered the connection between intermolecular forces and radiation processes.\cite{Lebedew1894,Derjaguin_2015} The theory linking optics and forces was subsequently established by Lifshitz and colleagues in their seminal papers.\cite{Lifshitz1956,Dzya} Initially, incorporating optical data across a broad frequency range seemed challenging for achieving highly accurate force calculations. However, Parsegian and Ninham demonstrated that a few oscillator models for the dielectric function could yield reasonably good agreement between theory and experimental forces.\cite{ParsegianNinham1969,NinhamParsegianWeiss1970,Richmond_1971_magnetic,Pars,Ninhb}
In our current study, we employed DFT to investigate the optical properties of two significant components, CaCO$_3$ and Al$_2$O$_3$, commonly found in diverse soil compositions. Our main objective was to investigate the influence of accurately describing the optical properties of these components on the formation of thin layers of water and ice on soil particles. This investigation carries significant scientific and engineering implications related to soil dynamics and related phenomena. Intriguingly, our findings reveal that the extrapolations made for low and high frequencies have minimal impact on the Hamaker constants and Lifshitz interactions. Importantly, all the extrapolations we investigated were reasonably accurate, as an inadequate extrapolation would lead to erroneous predictions for Casimir-Lifshitz forces and Hamaker constants, as evidenced by the divergent outcomes for the two materials. However, it is essential to acknowledge that the accuracy of these conclusions may vary for other systems, especially those characterized by crossings of different $\varepsilon(i\xi)$ functions at specific frequencies\,\cite{Elbaum,PhysRevB.89.201407} or involving metallic components.\cite{Bost2000,MathiasRizwanHarshanIverClasSashaPRB2023}
This work, along with prior research, establishes a direct link between optics derived from DFT and dispersion forces and their associated energies. We analyze these interactions and their impact on the formation of water and ice layers on soil particles in contact with moisture. Our findings reveal the previously unrecognized significance of this phenomenon in assessing soil water content, as highlighted by Lebron {\it{et al..}}\,\cite{LebronRobinsonGoldbergLeschCalcite2004} Furthermore, our predictions hold substantial implications for future models concerning frost heave,\cite{Rempel2007_frostheave} as well as related effects such as cement storage and degradation in moist environments. \,\cite{WEBER199759}

\begin{acknowledgments}
The authors thank the "ENSEMBLE3 - Centre of Excellence for nanophotonics, advanced materials and novel crystal growth-based technologies" project (GA No. MAB/2020/14) carried out within the International Research Agendas programme of the Foundation for Polish Science co-financed by the European Union under the European Regional Development Fund and the European Union's Horizon 2020 research and innovation programme Teaming for Excellence (GA. No. 857543) for support of this work. 
We acknowledge also the financial support from the European Union's Horizon 2020 research and innovation programme, grant agreements No.~869815 and No.~101058694 and 
the Research Council of Norway (Project No.~250346).
We acknowledge access to high-performance computing resources via NAISS, provided by NSC and PDC. All DFT calculations 
were performed at KTH Royal Institute of Technology (Sweden).
We finally acknowledge constructive discussions on related topics, over the last decade, with Dr. Kristian Berland. 
\end{acknowledgments}

\bibliography{PRB_Ensemble}

\begin{thebibliography}{53}%
\makeatletter
\providecommand \@ifxundefined [1]{%
 \@ifx{#1\undefined}
}%
\providecommand \@ifnum [1]{%
 \ifnum #1\expandafter \@firstoftwo
 \else \expandafter \@secondoftwo
 \fi
}%
\providecommand \@ifx [1]{%
 \ifx #1\expandafter \@firstoftwo
 \else \expandafter \@secondoftwo
 \fi
}%
\providecommand \natexlab [1]{#1}%
\providecommand \enquote  [1]{``#1''}%
\providecommand \bibnamefont  [1]{#1}%
\providecommand \bibfnamefont [1]{#1}%
\providecommand \citenamefont [1]{#1}%
\providecommand \href@noop [0]{\@secondoftwo}%
\providecommand \href [0]{\begingroup \@sanitize@url \@href}%
\providecommand \@href[1]{\@@startlink{#1}\@@href}%
\providecommand \@@href[1]{\endgroup#1\@@endlink}%
\providecommand \@sanitize@url [0]{\catcode `\\12\catcode `\$12\catcode `\&12\catcode `\#12\catcode `\^12\catcode `\_12\catcode `\%12\relax}%
\providecommand \@@startlink[1]{}%
\providecommand \@@endlink[0]{}%
\providecommand \url  [0]{\begingroup\@sanitize@url \@url }%
\providecommand \@url [1]{\endgroup\@href {#1}{\urlprefix }}%
\providecommand \urlprefix  [0]{URL }%
\providecommand \Eprint [0]{\href }%
\providecommand \doibase [0]{http://dx.doi.org/}%
\providecommand \selectlanguage [0]{\@gobble}%
\providecommand \bibinfo  [0]{\@secondoftwo}%
\providecommand \bibfield  [0]{\@secondoftwo}%
\providecommand \translation [1]{[#1]}%
\providecommand \BibitemOpen [0]{}%
\providecommand \bibitemStop [0]{}%
\providecommand \bibitemNoStop [0]{.\EOS\space}%
\providecommand \EOS [0]{\spacefactor3000\relax}%
\providecommand \BibitemShut  [1]{\csname bibitem#1\endcsname}%
\let\auto@bib@innerbib\@empty
\bibitem [{\citenamefont {Jacob}\ \emph {et~al.}(2012)\citenamefont {Jacob}, \citenamefont {Wahr}, \citenamefont {Pfeffer},\ and\ \citenamefont {Swenson}}]{JacobGlaciers2012}%
  \BibitemOpen
  \bibfield  {author} {\bibinfo {author} {\bibfnamefont {T.}~\bibnamefont {Jacob}}, \bibinfo {author} {\bibfnamefont {J.}~\bibnamefont {Wahr}}, \bibinfo {author} {\bibfnamefont {T.}~\bibnamefont {Pfeffer}}, \ and\ \bibinfo {author} {\bibfnamefont {S.}~\bibnamefont {Swenson}},\ }\bibfield  {title} {\enquote {\bibinfo {title} {Recent contributions of glaciers and ice caps to sea level rise},}\ }\href@noop {} {\bibfield  {journal} {\bibinfo  {journal} {Nature}\ }\textbf {\bibinfo {volume} {482}},\ \bibinfo {pages} {514--518} (\bibinfo {year} {2012})}\BibitemShut {NoStop}%
\bibitem [{\citenamefont {Lannuzel}\ \emph {et~al.}(2020)\citenamefont {Lannuzel}, \citenamefont {Tedesco},\ and\ \citenamefont {van Leeuwe~et al.}}]{LannuzelNatClimCha2020}%
  \BibitemOpen
  \bibfield  {author} {\bibinfo {author} {\bibfnamefont {D.}~\bibnamefont {Lannuzel}}, \bibinfo {author} {\bibfnamefont {L.}~\bibnamefont {Tedesco}}, \ and\ \bibinfo {author} {\bibfnamefont {M.}~\bibnamefont {van Leeuwe~et al.}},\ }\bibfield  {title} {\enquote {\bibinfo {title} {The future of arctic sea-ice biogeochemistry and ice-associated ecosystems},}\ }\href@noop {} {\bibfield  {journal} {\bibinfo  {journal} {Nature Climate Change}\ }\textbf {\bibinfo {volume} {10}},\ \bibinfo {pages} {983--992} (\bibinfo {year} {2020})}\BibitemShut {NoStop}%
\bibitem [{\citenamefont {Weber}\ and\ \citenamefont {Reinhardt}(1997)}]{WEBER199759}%
  \BibitemOpen
  \bibfield  {author} {\bibinfo {author} {\bibfnamefont {S.}~\bibnamefont {Weber}}\ and\ \bibinfo {author} {\bibfnamefont {H.~W.}\ \bibnamefont {Reinhardt}},\ }\bibfield  {title} {\enquote {\bibinfo {title} {A new generation of high performance concrete: Concrete with autogenous curing},}\ }\href@noop {} {\bibfield  {journal} {\bibinfo  {journal} {Advanced Cement Based Materials}\ }\textbf {\bibinfo {volume} {6}},\ \bibinfo {pages} {59--68} (\bibinfo {year} {1997})}\BibitemShut {NoStop}%
\bibitem [{\citenamefont {Hamada}\ \emph {et~al.}(2023)\citenamefont {Hamada}, \citenamefont {Abdulhaleem}, \citenamefont {Majdi}, \citenamefont {{Al Jawahery}}, \citenamefont {{Skariah Thomas}},\ and\ \citenamefont {Yousif}}]{HAMADA2023}%
  \BibitemOpen
  \bibfield  {author} {\bibinfo {author} {\bibfnamefont {H.~M.}\ \bibnamefont {Hamada}}, \bibinfo {author} {\bibfnamefont {K.~N.}\ \bibnamefont {Abdulhaleem}}, \bibinfo {author} {\bibfnamefont {A.}~\bibnamefont {Majdi}}, \bibinfo {author} {\bibfnamefont {M.~S.}\ \bibnamefont {{Al Jawahery}}}, \bibinfo {author} {\bibfnamefont {B.}~\bibnamefont {{Skariah Thomas}}}, \ and\ \bibinfo {author} {\bibfnamefont {S.~T.}\ \bibnamefont {Yousif}},\ }\bibfield  {title} {\enquote {\bibinfo {title} {Effect of wastewater as sustainable concrete material on concrete performance: A critical review},}\ }\href@noop {} {\bibfield  {journal} {\bibinfo  {journal} {Materials Today: Proceedings}\ } (\bibinfo {year} {2023})}\BibitemShut {NoStop}%
\bibitem [{\citenamefont {Tiessen}\ \emph {et~al.}(1994)\citenamefont {Tiessen}, \citenamefont {Cuevas},\ and\ \citenamefont {Chacon}}]{TiessenNature1994}%
  \BibitemOpen
  \bibfield  {author} {\bibinfo {author} {\bibfnamefont {H}~\bibnamefont {Tiessen}}, \bibinfo {author} {\bibfnamefont {E.}~\bibnamefont {Cuevas}}, \ and\ \bibinfo {author} {\bibfnamefont {P.}~\bibnamefont {Chacon}},\ }\bibfield  {title} {\enquote {\bibinfo {title} {The role of soil organic matter in sustaining soil fertility},}\ }\href@noop {} {\bibfield  {journal} {\bibinfo  {journal} {Nature}\ }\textbf {\bibinfo {volume} {371}},\ \bibinfo {pages} {783--785} (\bibinfo {year} {1994})}\BibitemShut {NoStop}%
\bibitem [{\citenamefont {Rempel}(2007)}]{Rempel2007_frostheave}%
  \BibitemOpen
  \bibfield  {author} {\bibinfo {author} {\bibfnamefont {A.~W.}\ \bibnamefont {Rempel}},\ }\bibfield  {title} {\enquote {\bibinfo {title} {Formation of ice lenses and frost heave},}\ }\href@noop {} {\bibfield  {journal} {\bibinfo  {journal} {Journal of Geophysical Research: Earth Surface}\ }\textbf {\bibinfo {volume} {112}} (\bibinfo {year} {2007})}\BibitemShut {NoStop}%
\bibitem [{\citenamefont {Luengo-M\'arquez}\ and\ \citenamefont {MacDowell}(2021)}]{LUENGOMARQUEZMacDowell2021}%
  \BibitemOpen
  \bibfield  {author} {\bibinfo {author} {\bibfnamefont {J.}~\bibnamefont {Luengo-M\'arquez}}\ and\ \bibinfo {author} {\bibfnamefont {L.~G.}\ \bibnamefont {MacDowell}},\ }\bibfield  {title} {\enquote {\bibinfo {title} {Lifshitz theory of wetting films at three phase coexistence: The case of ice nucleation on silver iodide (agi)},}\ }\href@noop {} {\bibfield  {journal} {\bibinfo  {journal} {J. Coll. Interf. Sci.}\ }\textbf {\bibinfo {volume} {590}},\ \bibinfo {pages} {527--538} (\bibinfo {year} {2021})}\BibitemShut {NoStop}%
\bibitem [{\citenamefont {Luengo-Marquez}\ \emph {et~al.}(2022)\citenamefont {Luengo-Marquez}, \citenamefont {Izquierdo-Ruiz},\ and\ \citenamefont {MacDowell}}]{luengo2022WaterIce}%
  \BibitemOpen
  \bibfield  {author} {\bibinfo {author} {\bibfnamefont {J.}~\bibnamefont {Luengo-Marquez}}, \bibinfo {author} {\bibfnamefont {F.}~\bibnamefont {Izquierdo-Ruiz}}, \ and\ \bibinfo {author} {\bibfnamefont {L.~G.}\ \bibnamefont {MacDowell}},\ }\bibfield  {title} {\enquote {\bibinfo {title} {Intermolecular forces at ice and water interfaces: Premelting, surface freezing, and regelation},}\ }\href@noop {} {\bibfield  {journal} {\bibinfo  {journal} {J. Chem. Phys.}\ }\textbf {\bibinfo {volume} {157}},\ \bibinfo {pages} {044704} (\bibinfo {year} {2022})}\BibitemShut {NoStop}%
\bibitem [{\citenamefont {Bostr\"om}\ \emph {et~al.}(2021)\citenamefont {Bostr\"om}, \citenamefont {Esteso}, \citenamefont {Fiedler}, \citenamefont {Brevik}, \citenamefont {Buhmann}, \citenamefont {Persson}, \citenamefont {Carretero-Palacios}, \citenamefont {Parsons},\ and\ \citenamefont {Corkery}}]{BostromEstesoFiedlerBrevikBuhmannPerssonCarreteroParsonsCorkery2021}%
  \BibitemOpen
  \bibfield  {author} {\bibinfo {author} {\bibfnamefont {M.}~\bibnamefont {Bostr\"om}}, \bibinfo {author} {\bibfnamefont {V.}~\bibnamefont {Esteso}}, \bibinfo {author} {\bibfnamefont {J.}~\bibnamefont {Fiedler}}, \bibinfo {author} {\bibfnamefont {I.}~\bibnamefont {Brevik}}, \bibinfo {author} {\bibfnamefont {S.~Y.}\ \bibnamefont {Buhmann}}, \bibinfo {author} {\bibfnamefont {C.}~\bibnamefont {Persson}}, \bibinfo {author} {\bibfnamefont {S.}~\bibnamefont {Carretero-Palacios}}, \bibinfo {author} {\bibfnamefont {D.~F.}\ \bibnamefont {Parsons}}, \ and\ \bibinfo {author} {\bibfnamefont {R.~W.}\ \bibnamefont {Corkery}},\ }\bibfield  {title} {\enquote {\bibinfo {title} {{Self-preserving ice layers on \ce{CO2} clathrate particles: implications for Enceladus, Pluto and similar ocean worlds}},}\ }\href@noop {} {\bibfield  {journal} {\bibinfo  {journal} {Astronomy and Astrophysics}\ }\textbf {\bibinfo {volume} {650}},\ \bibinfo {pages} {A54} (\bibinfo {year} {2021})}\BibitemShut {NoStop}%
\bibitem [{\citenamefont {Waite}\ \emph {et~al.}(2006)\citenamefont {Waite}, \citenamefont {Combi}, \citenamefont {Ip}, \citenamefont {Cravens}, \citenamefont {McNutt}, \citenamefont {Kasprzak}, \citenamefont {Yelle}, \citenamefont {Luhmann}, \citenamefont {Niemann}, \citenamefont {Gell}, \citenamefont {Magee}, \citenamefont {Fletcher}, \citenamefont {Lunine},\ and\ \citenamefont {Tseng}}]{Waite2006}%
  \BibitemOpen
  \bibfield  {author} {\bibinfo {author} {\bibfnamefont {J.~H.}\ \bibnamefont {Waite}}, \bibinfo {author} {\bibfnamefont {M.~R.}\ \bibnamefont {Combi}}, \bibinfo {author} {\bibfnamefont {W.~H.}\ \bibnamefont {Ip}}, \bibinfo {author} {\bibfnamefont {T.~E.}\ \bibnamefont {Cravens}}, \bibinfo {author} {\bibfnamefont {R.~L.}\ \bibnamefont {McNutt}}, \bibinfo {author} {\bibfnamefont {W.}~\bibnamefont {Kasprzak}}, \bibinfo {author} {\bibfnamefont {R.}~\bibnamefont {Yelle}}, \bibinfo {author} {\bibfnamefont {J.}~\bibnamefont {Luhmann}}, \bibinfo {author} {\bibfnamefont {H.}~\bibnamefont {Niemann}}, \bibinfo {author} {\bibfnamefont {D.}~\bibnamefont {Gell}}, \bibinfo {author} {\bibfnamefont {B.}~\bibnamefont {Magee}}, \bibinfo {author} {\bibfnamefont {G.}~\bibnamefont {Fletcher}}, \bibinfo {author} {\bibfnamefont {G.}~\bibnamefont {Lunine}}, \ and\ \bibinfo {author} {\bibfnamefont {W.~L}\ \bibnamefont {Tseng}},\ }\bibfield  {title} {\enquote {\bibinfo {title} {{Cassini ion and neutral mass spectrometer: Enceladus
  plume composition and structure}},}\ }\href@noop {} {\bibfield  {journal} {\bibinfo  {journal} {Science}\ }\textbf {\bibinfo {volume} {311}},\ \bibinfo {pages} {1419--1422} (\bibinfo {year} {2006})}\BibitemShut {NoStop}%
\bibitem [{\citenamefont {Muhs}(2001)}]{Muhs2001}%
  \BibitemOpen
  \bibfield  {author} {\bibinfo {author} {\bibfnamefont {D.~R.}\ \bibnamefont {Muhs}},\ }\bibfield  {title} {\enquote {\bibinfo {title} {Evolution of soils on quaternary reef terraces of barbados, west indies},}\ }\href@noop {} {\bibfield  {journal} {\bibinfo  {journal} {Quaternary Research}\ }\textbf {\bibinfo {volume} {56}},\ \bibinfo {pages} {66--78} (\bibinfo {year} {2001})}\BibitemShut {NoStop}%
\bibitem [{\citenamefont {Nordt}\ and\ \citenamefont {Driese}(2010)}]{Nordt2010}%
  \BibitemOpen
  \bibfield  {author} {\bibinfo {author} {\bibfnamefont {L.~C.}\ \bibnamefont {Nordt}}\ and\ \bibinfo {author} {\bibfnamefont {S.~G.}\ \bibnamefont {Driese}},\ }\bibfield  {title} {\enquote {\bibinfo {title} {A modern soil characterization approach to reconstructing physical and chemical properties of paleo-vertisols},}\ }\href@noop {} {\bibfield  {journal} {\bibinfo  {journal} {American Journal of Science}\ }\textbf {\bibinfo {volume} {310}},\ \bibinfo {pages} {37--64} (\bibinfo {year} {2010})}\BibitemShut {NoStop}%
\bibitem [{\citenamefont {Parsegian}\ and\ \citenamefont {Ninham}(1969)}]{ParsegianNinham1969}%
  \BibitemOpen
  \bibfield  {author} {\bibinfo {author} {\bibfnamefont {V.~A.}\ \bibnamefont {Parsegian}}\ and\ \bibinfo {author} {\bibfnamefont {B.~W.}\ \bibnamefont {Ninham}},\ }\bibfield  {title} {\enquote {\bibinfo {title} {{Application of the Lifshitz theory to the calculation of Van der Waals forces across thin lipid films}},}\ }\href@noop {} {\bibfield  {journal} {\bibinfo  {journal} {Nature}\ }\textbf {\bibinfo {volume} {224}},\ \bibinfo {pages} {1197--1198} (\bibinfo {year} {1969})}\BibitemShut {NoStop}%
\bibitem [{\citenamefont {Lebron}\ \emph {et~al.}(2004)\citenamefont {Lebron}, \citenamefont {Robinson}, \citenamefont {Goldberg},\ and\ \citenamefont {Lesch}}]{LebronRobinsonGoldbergLeschCalcite2004}%
  \BibitemOpen
  \bibfield  {author} {\bibinfo {author} {\bibfnamefont {I.}~\bibnamefont {Lebron}}, \bibinfo {author} {\bibfnamefont {D.~A.}\ \bibnamefont {Robinson}}, \bibinfo {author} {\bibfnamefont {S.}~\bibnamefont {Goldberg}}, \ and\ \bibinfo {author} {\bibfnamefont {S.~M.}\ \bibnamefont {Lesch}},\ }\bibfield  {title} {\enquote {\bibinfo {title} {The dielectric permittivity of calcite and arid zone soils with carbonate minerals},}\ }\href@noop {} {\bibfield  {journal} {\bibinfo  {journal} {Soil Science Society of America Journal}\ }\textbf {\bibinfo {volume} {68}},\ \bibinfo {pages} {1549--1559} (\bibinfo {year} {2004})}\BibitemShut {NoStop}%
\bibitem [{\citenamefont {Lifshitz}(1956)}]{Lifshitz1956}%
  \BibitemOpen
  \bibfield  {author} {\bibinfo {author} {\bibfnamefont {E.~M.}\ \bibnamefont {Lifshitz}},\ }\bibfield  {title} {\enquote {\bibinfo {title} {The theory of molecular attractive forces between solids},}\ }\href@noop {} {\bibfield  {journal} {\bibinfo  {journal} {Sov. Phys. JETP}\ }\textbf {\bibinfo {volume} {2}},\ \bibinfo {pages} {73} (\bibinfo {year} {1956})}\BibitemShut {NoStop}%
\bibitem [{\citenamefont {Dzyaloshinskii}\ \emph {et~al.}(1961)\citenamefont {Dzyaloshinskii}, \citenamefont {Lifshitz},\ and\ \citenamefont {Pitaevskii}}]{Dzya}%
  \BibitemOpen
  \bibfield  {author} {\bibinfo {author} {\bibfnamefont {I.E.}\ \bibnamefont {Dzyaloshinskii}}, \bibinfo {author} {\bibfnamefont {E.M.}\ \bibnamefont {Lifshitz}}, \ and\ \bibinfo {author} {\bibfnamefont {L.P.}\ \bibnamefont {Pitaevskii}},\ }\bibfield  {title} {\enquote {\bibinfo {title} {{The general theory of van der Waals forces}},}\ }\href@noop {} {\bibfield  {journal} {\bibinfo  {journal} {Adv. Phys.}\ }\textbf {\bibinfo {volume} {10}},\ \bibinfo {pages} {165--209} (\bibinfo {year} {1961})}\BibitemShut {NoStop}%
\bibitem [{\citenamefont {van Kampen}\ \emph {et~al.}(1968)\citenamefont {van Kampen}, \citenamefont {Nijboer},\ and\ \citenamefont {Schram}}]{KampenNijboerSchram1968}%
  \BibitemOpen
  \bibfield  {author} {\bibinfo {author} {\bibfnamefont {N.~G.}\ \bibnamefont {van Kampen}}, \bibinfo {author} {\bibfnamefont {B.~R.~A.}\ \bibnamefont {Nijboer}}, \ and\ \bibinfo {author} {\bibfnamefont {K.}~\bibnamefont {Schram}},\ }\bibfield  {title} {\enquote {\bibinfo {title} {{On the macroscopic theory of Van der Waals forces}},}\ }\href@noop {} {\bibfield  {journal} {\bibinfo  {journal} {Phys. Lett.}\ }\textbf {\bibinfo {volume} {A26}},\ \bibinfo {pages} {307} (\bibinfo {year} {1968})}\BibitemShut {NoStop}%
\bibitem [{\citenamefont {Ninham}\ \emph {et~al.}(1970)\citenamefont {Ninham}, \citenamefont {Parsegian},\ and\ \citenamefont {Weiss}}]{NinhamParsegianWeiss1970}%
  \BibitemOpen
  \bibfield  {author} {\bibinfo {author} {\bibfnamefont {B.~W.}\ \bibnamefont {Ninham}}, \bibinfo {author} {\bibfnamefont {V.~A.}\ \bibnamefont {Parsegian}}, \ and\ \bibinfo {author} {\bibfnamefont {G.~H.}\ \bibnamefont {Weiss}},\ }\bibfield  {title} {\enquote {\bibinfo {title} {{On the macroscopic theory of temperature dependent Van der Waals forces}},}\ }\href@noop {} {\bibfield  {journal} {\bibinfo  {journal} {J. Stat. Phys.}\ }\textbf {\bibinfo {volume} {2}},\ \bibinfo {pages} {323} (\bibinfo {year} {1970})}\BibitemShut {NoStop}%
\bibitem [{\citenamefont {Richmond}\ and\ \citenamefont {Ninham}(1971)}]{Richmond_1971_magnetic}%
  \BibitemOpen
  \bibfield  {author} {\bibinfo {author} {\bibfnamefont {P.}~\bibnamefont {Richmond}}\ and\ \bibinfo {author} {\bibfnamefont {B.~W.}\ \bibnamefont {Ninham}},\ }\bibfield  {title} {\enquote {\bibinfo {title} {{A note on the extension of the Lifshitz theory of van der Waals forces to magnetic media}},}\ }\href@noop {} {\bibfield  {journal} {\bibinfo  {journal} {J. Phys. C: Solid St. Phys}\ }\textbf {\bibinfo {volume} {4}},\ \bibinfo {pages} {1988} (\bibinfo {year} {1971})}\BibitemShut {NoStop}%
\bibitem [{\citenamefont {Barash}\ and\ \citenamefont {Ginzburg}(1975)}]{BarashGinzburg1975}%
  \BibitemOpen
  \bibfield  {author} {\bibinfo {author} {\bibfnamefont {Yu.~S.}\ \bibnamefont {Barash}}\ and\ \bibinfo {author} {\bibfnamefont {V.~L.}\ \bibnamefont {Ginzburg}},\ }\bibfield  {title} {\enquote {\bibinfo {title} {{Electromagnetic fluctuations in matter and molecular (Van-der-Waals) forces between them}},}\ }\href@noop {} {\bibfield  {journal} {\bibinfo  {journal} {Sov. Phys.-Usp.}\ }\textbf {\bibinfo {volume} {18}},\ \bibinfo {pages} {305} (\bibinfo {year} {1975})}\BibitemShut {NoStop}%
\bibitem [{\citenamefont {Landau}\ and\ \citenamefont {Lifshitz}(2013)}]{landau2013statistical}%
  \BibitemOpen
  \bibfield  {author} {\bibinfo {author} {\bibfnamefont {L.~D.}\ \bibnamefont {Landau}}\ and\ \bibinfo {author} {\bibfnamefont {E.~M.}\ \bibnamefont {Lifshitz}},\ }\href@noop {} {\emph {\bibinfo {title} {Statistical Physics, Third Edition, Part 1: Volume 5}}}\ (\bibinfo  {publisher} {Elsevier},\ \bibinfo {year} {2013})\BibitemShut {NoStop}%
\bibitem [{\citenamefont {Haydon}\ and\ \citenamefont {Taylor}(1968)}]{HaydonTaylor1968}%
  \BibitemOpen
  \bibfield  {author} {\bibinfo {author} {\bibfnamefont {D.~A.}\ \bibnamefont {Haydon}}\ and\ \bibinfo {author} {\bibfnamefont {J.~L.}\ \bibnamefont {Taylor}},\ }\bibfield  {title} {\enquote {\bibinfo {title} {{Contact Angles for Thin Lipid Films and the Determination of London-van der Waals Forces}},}\ }\href@noop {} {\bibfield  {journal} {\bibinfo  {journal} {Nature}\ }\textbf {\bibinfo {volume} {217}},\ \bibinfo {pages} {739--740} (\bibinfo {year} {1968})}\BibitemShut {NoStop}%
\bibitem [{\citenamefont {Kresse}\ and\ \citenamefont {Joubert}(1999)}]{VASP1999}%
  \BibitemOpen
  \bibfield  {author} {\bibinfo {author} {\bibfnamefont {G.}~\bibnamefont {Kresse}}\ and\ \bibinfo {author} {\bibfnamefont {D.}~\bibnamefont {Joubert}},\ }\bibfield  {title} {\enquote {\bibinfo {title} {From ultrasoft pseudopotentials to the projector augmented-wave method},}\ }\href@noop {} {\bibfield  {journal} {\bibinfo  {journal} {Phys. Rev. B}\ }\textbf {\bibinfo {volume} {59}},\ \bibinfo {pages} {1758--1775} (\bibinfo {year} {1999})}\BibitemShut {NoStop}%
\bibitem [{\citenamefont {Perdew}\ \emph {et~al.}(2008)\citenamefont {Perdew}, \citenamefont {Ruzsinszky}, \citenamefont {Csonka}, \citenamefont {Vydrov}, \citenamefont {Scuseria}, \citenamefont {Constantin}, \citenamefont {Zhou},\ and\ \citenamefont {Burke}}]{PBEsol2008}%
  \BibitemOpen
  \bibfield  {author} {\bibinfo {author} {\bibfnamefont {J.~P.}\ \bibnamefont {Perdew}}, \bibinfo {author} {\bibfnamefont {A.}~\bibnamefont {Ruzsinszky}}, \bibinfo {author} {\bibfnamefont {G.}~\bibnamefont {Csonka}}, \bibinfo {author} {\bibfnamefont {O.~A.}\ \bibnamefont {Vydrov}}, \bibinfo {author} {\bibfnamefont {G.~E.}\ \bibnamefont {Scuseria}}, \bibinfo {author} {\bibfnamefont {L.~A.}\ \bibnamefont {Constantin}}, \bibinfo {author} {\bibfnamefont {X.}~\bibnamefont {Zhou}}, \ and\ \bibinfo {author} {\bibfnamefont {K.}~\bibnamefont {Burke}},\ }\bibfield  {title} {\enquote {\bibinfo {title} {Restoring the density-gradient expansion for exchange in solids and surfaces},}\ }\href@noop {} {\bibfield  {journal} {\bibinfo  {journal} {Phys. Rev. Lett.}\ }\textbf {\bibinfo {volume} {100}},\ \bibinfo {pages} {136406} (\bibinfo {year} {2008})}\BibitemShut {NoStop}%
\bibitem [{\citenamefont {Gao}\ \emph {et~al.}(2021)\citenamefont {Gao}, \citenamefont {Wu}, \citenamefont {Persson},\ and\ \citenamefont {Wang}}]{Gao2021}%
  \BibitemOpen
  \bibfield  {author} {\bibinfo {author} {\bibfnamefont {J.}~\bibnamefont {Gao}}, \bibinfo {author} {\bibfnamefont {Q.}~\bibnamefont {Wu}}, \bibinfo {author} {\bibfnamefont {C.}~\bibnamefont {Persson}}, \ and\ \bibinfo {author} {\bibfnamefont {Z.}~\bibnamefont {Wang}},\ }\bibfield  {title} {\enquote {\bibinfo {title} {{Irvsp: To obtain irreducible representations of electronic states in the VASP}},}\ }\href@noop {} {\bibfield  {journal} {\bibinfo  {journal} {Comput. Phys. Commun.}\ }\textbf {\bibinfo {volume} {261}},\ \bibinfo {pages} {107760} (\bibinfo {year} {2021})}\BibitemShut {NoStop}%
\bibitem [{\citenamefont {Crovetto}\ \emph {et~al.}(2016)\citenamefont {Crovetto}, \citenamefont {Chen}, \citenamefont {Ettlinger}, \citenamefont {Cazzaniga}, \citenamefont {Schou}, \citenamefont {Persson},\ and\ \citenamefont {Hansen}}]{CROVETTO2016}%
  \BibitemOpen
  \bibfield  {author} {\bibinfo {author} {\bibfnamefont {A.}~\bibnamefont {Crovetto}}, \bibinfo {author} {\bibfnamefont {R.}~\bibnamefont {Chen}}, \bibinfo {author} {\bibfnamefont {R.~B.}\ \bibnamefont {Ettlinger}}, \bibinfo {author} {\bibfnamefont {A.~C.}\ \bibnamefont {Cazzaniga}}, \bibinfo {author} {\bibfnamefont {J.}~\bibnamefont {Schou}}, \bibinfo {author} {\bibfnamefont {C.}~\bibnamefont {Persson}}, \ and\ \bibinfo {author} {\bibfnamefont {O.}~\bibnamefont {Hansen}},\ }\bibfield  {title} {\enquote {\bibinfo {title} {{Dielectric function and double absorption onset of monoclinic CuSnS3: Origin of experimental features explained by first-principles calculations}},}\ }\href@noop {} {\bibfield  {journal} {\bibinfo  {journal} {Sol. Energy Mater. Sol. Cells}\ }\textbf {\bibinfo {volume} {154}},\ \bibinfo {pages} {121--129} (\bibinfo {year} {2016})}\BibitemShut {NoStop}%
\bibitem [{\citenamefont {Persson}\ and\ \citenamefont {Lindefelt}(1999)}]{Persson1999}%
  \BibitemOpen
  \bibfield  {author} {\bibinfo {author} {\bibfnamefont {C.}~\bibnamefont {Persson}}\ and\ \bibinfo {author} {\bibfnamefont {U.}~\bibnamefont {Lindefelt}},\ }\bibfield  {title} {\enquote {\bibinfo {title} {{Dependence of energy gaps and effective masses on atomic positions in hexagonal SiC}},}\ }\href@noop {} {\bibfield  {journal} {\bibinfo  {journal} {J. App. Phys.}\ }\textbf {\bibinfo {volume} {86}},\ \bibinfo {pages} {5036--5039} (\bibinfo {year} {1999})}\BibitemShut {NoStop}%
\bibitem [{\citenamefont {d'Amour}\ \emph {et~al.}(1978)\citenamefont {d'Amour}, \citenamefont {Schiferl}, \citenamefont {Denner}, \citenamefont {Schulz},\ and\ \citenamefont {Holzapfel}}]{dAmour1978}%
  \BibitemOpen
  \bibfield  {author} {\bibinfo {author} {\bibfnamefont {H.}~\bibnamefont {d'Amour}}, \bibinfo {author} {\bibfnamefont {D.}~\bibnamefont {Schiferl}}, \bibinfo {author} {\bibfnamefont {W.}~\bibnamefont {Denner}}, \bibinfo {author} {\bibfnamefont {Heinz}\ \bibnamefont {Schulz}}, \ and\ \bibinfo {author} {\bibfnamefont {W.~B.}\ \bibnamefont {Holzapfel}},\ }\bibfield  {title} {\enquote {\bibinfo {title} {High-pressure single-crystal structure determinations for ruby up to 90 kbar using an automatic diffractometer},}\ }\href@noop {} {\bibfield  {journal} {\bibinfo  {journal} {J. Appl. Phys.}\ }\textbf {\bibinfo {volume} {49}},\ \bibinfo {pages} {4411--4416} (\bibinfo {year} {1978})}\BibitemShut {NoStop}%
\bibitem [{\citenamefont {Lucht}\ \emph {et~al.}(2003)\citenamefont {Lucht}, \citenamefont {Lerche}, \citenamefont {Wille}, \citenamefont {Shvyd'ko}, \citenamefont {R{\"{u}}ter}, \citenamefont {Gerdau},\ and\ \citenamefont {Becker}}]{Lucht2003}%
  \BibitemOpen
  \bibfield  {author} {\bibinfo {author} {\bibfnamefont {M.}~\bibnamefont {Lucht}}, \bibinfo {author} {\bibfnamefont {M.}~\bibnamefont {Lerche}}, \bibinfo {author} {\bibfnamefont {H.-C.}\ \bibnamefont {Wille}}, \bibinfo {author} {\bibfnamefont {Yu.~V.}\ \bibnamefont {Shvyd'ko}}, \bibinfo {author} {\bibfnamefont {H.~D.}\ \bibnamefont {R{\"{u}}ter}}, \bibinfo {author} {\bibfnamefont {E.}~\bibnamefont {Gerdau}}, \ and\ \bibinfo {author} {\bibfnamefont {P.}~\bibnamefont {Becker}},\ }\bibfield  {title} {\enquote {\bibinfo {title} {{Precise measurement of the lattice parameters of {$\alpha$}-Al${\sb 2}$O${\sb 3}$ in the temperature range 4.5{--}250K using the M{\"{o}}ssbauer wavelength standard}},}\ }\href@noop {} {\bibfield  {journal} {\bibinfo  {journal} {J. Appl. Crystallogr.}\ }\textbf {\bibinfo {volume} {36}},\ \bibinfo {pages} {1075--1081} (\bibinfo {year} {2003})}\BibitemShut {NoStop}%
\bibitem [{\citenamefont {Oetzel}\ and\ \citenamefont {Heger}(1999)}]{Oetzel1999}%
  \BibitemOpen
  \bibfield  {author} {\bibinfo {author} {\bibfnamefont {M.}~\bibnamefont {Oetzel}}\ and\ \bibinfo {author} {\bibfnamefont {G.}~\bibnamefont {Heger}},\ }\bibfield  {title} {\enquote {\bibinfo {title} {{Laboratory X-ray powder diffraction: a comparison of different geometries with special attention to the usage of the Cu {\it K}{$\alpha$} doublet}},}\ }\href@noop {} {\bibfield  {journal} {\bibinfo  {journal} {J. Appl. Cryst.}\ }\textbf {\bibinfo {volume} {32}},\ \bibinfo {pages} {799--807} (\bibinfo {year} {1999})}\BibitemShut {NoStop}%
\bibitem [{\citenamefont {French}(1990)}]{French1990}%
  \BibitemOpen
  \bibfield  {author} {\bibinfo {author} {\bibfnamefont {R.~H.}\ \bibnamefont {French}},\ }\bibfield  {title} {\enquote {\bibinfo {title} {{Electronic Band Structure of Al2O3, with Comparison to Alon and AIN}},}\ }\href@noop {} {\bibfield  {journal} {\bibinfo  {journal} {J. Am. Ceram. Soc.}\ }\textbf {\bibinfo {volume} {73}},\ \bibinfo {pages} {477--489} (\bibinfo {year} {1990})}\BibitemShut {NoStop}%
\bibitem [{\citenamefont {Bortz}\ and\ \citenamefont {French}(1989)}]{Bortz1989}%
  \BibitemOpen
  \bibfield  {author} {\bibinfo {author} {\bibfnamefont {M.~L.}\ \bibnamefont {Bortz}}\ and\ \bibinfo {author} {\bibfnamefont {R.~H.}\ \bibnamefont {French}},\ }\bibfield  {title} {\enquote {\bibinfo {title} {Optical reflectivity measurements using a laser plasma light source},}\ }\href@noop {} {\bibfield  {journal} {\bibinfo  {journal} {Appl. Phys. Lett.}\ }\textbf {\bibinfo {volume} {55}},\ \bibinfo {pages} {1955--1957} (\bibinfo {year} {1989})}\BibitemShut {NoStop}%
\bibitem [{\citenamefont {Maslen}\ \emph {et~al.}(1995)\citenamefont {Maslen}, \citenamefont {Streltsov}, \citenamefont {Streltsova},\ and\ \citenamefont {Ishizawa}}]{Maslen1995}%
  \BibitemOpen
  \bibfield  {author} {\bibinfo {author} {\bibfnamefont {E.~N.}\ \bibnamefont {Maslen}}, \bibinfo {author} {\bibfnamefont {V.~A.}\ \bibnamefont {Streltsov}}, \bibinfo {author} {\bibfnamefont {N.~R.}\ \bibnamefont {Streltsova}}, \ and\ \bibinfo {author} {\bibfnamefont {N.}~\bibnamefont {Ishizawa}},\ }\bibfield  {title} {\enquote {\bibinfo {title} {{Electron density and optical anisotropy in rhombohedral carbonates. III. Synchrotron X-ray studies of CaCO${\sb 3}$, MgCO${\sb 3}$ and MnCO${\sb 3}$}},}\ }\href@noop {} {\bibfield  {journal} {\bibinfo  {journal} {Acta Cryst. B}\ }\textbf {\bibinfo {volume} {51}},\ \bibinfo {pages} {929--939} (\bibinfo {year} {1995})}\BibitemShut {NoStop}%
\bibitem [{\citenamefont {Graf}(1961)}]{Graf1961}%
  \BibitemOpen
  \bibfield  {author} {\bibinfo {author} {\bibfnamefont {D.~L.}\ \bibnamefont {Graf}},\ }\bibfield  {title} {\enquote {\bibinfo {title} {{Crystallographic tables for the rhombohedral carbonates}},}\ }\href@noop {} {\bibfield  {journal} {\bibinfo  {journal} {Am. Mineral.}\ }\textbf {\bibinfo {volume} {46}},\ \bibinfo {pages} {1283--1316} (\bibinfo {year} {1961})}\BibitemShut {NoStop}%
\bibitem [{\citenamefont {Karunadasa}\ \emph {et~al.}(2019)\citenamefont {Karunadasa}, \citenamefont {Manoratne}, \citenamefont {Pitawala},\ and\ \citenamefont {Rajapakse}}]{Karunadasa2019}%
  \BibitemOpen
  \bibfield  {author} {\bibinfo {author} {\bibfnamefont {K.S.P.}\ \bibnamefont {Karunadasa}}, \bibinfo {author} {\bibfnamefont {C.H.}\ \bibnamefont {Manoratne}}, \bibinfo {author} {\bibfnamefont {H.M.T.G.A.}\ \bibnamefont {Pitawala}}, \ and\ \bibinfo {author} {\bibfnamefont {R.M.G.}\ \bibnamefont {Rajapakse}},\ }\bibfield  {title} {\enquote {\bibinfo {title} {Thermal decomposition of calcium carbonate (calcite polymorph) as examined by in-situ high-temperature x-ray powder diffraction},}\ }\href@noop {} {\bibfield  {journal} {\bibinfo  {journal} {J. Phys. Chem. Solids}\ }\textbf {\bibinfo {volume} {134}},\ \bibinfo {pages} {21--28} (\bibinfo {year} {2019})}\BibitemShut {NoStop}%
\bibitem [{\citenamefont {Baer}\ and\ \citenamefont {Blanchard}(1993)}]{Baer1993}%
  \BibitemOpen
  \bibfield  {author} {\bibinfo {author} {\bibfnamefont {D.R.}\ \bibnamefont {Baer}}\ and\ \bibinfo {author} {\bibfnamefont {D.~L.}\ \bibnamefont {Blanchard}},\ }\bibfield  {title} {\enquote {\bibinfo {title} {Studies of the calcite cleavage surface for comparison with calculation},}\ }\href@noop {} {\bibfield  {journal} {\bibinfo  {journal} {Appl. Surf. Sci.}\ }\textbf {\bibinfo {volume} {72}},\ \bibinfo {pages} {295--300} (\bibinfo {year} {1993})}\BibitemShut {NoStop}%
\bibitem [{\citenamefont {Vos}\ \emph {et~al.}(2015)\citenamefont {Vos}, \citenamefont {Marmitt}, \citenamefont {Finkelstein},\ and\ \citenamefont {Moreh}}]{Vos2015}%
  \BibitemOpen
  \bibfield  {author} {\bibinfo {author} {\bibfnamefont {M.}~\bibnamefont {Vos}}, \bibinfo {author} {\bibfnamefont {G.~G.}\ \bibnamefont {Marmitt}}, \bibinfo {author} {\bibfnamefont {Y.}~\bibnamefont {Finkelstein}}, \ and\ \bibinfo {author} {\bibfnamefont {R.}~\bibnamefont {Moreh}},\ }\bibfield  {title} {\enquote {\bibinfo {title} {Determining the band gap and mean kinetic energy of atoms from reflection electron energy loss spectra},}\ }\href@noop {} {\bibfield  {journal} {\bibinfo  {journal} {J. Chem. Phys.}\ }\textbf {\bibinfo {volume} {143}},\ \bibinfo {pages} {104203} (\bibinfo {year} {2015})}\BibitemShut {NoStop}%
\bibitem [{\citenamefont {at~[url will be inserted by publisher] for tables of the~irreducible representations}\ and\ \citenamefont {detailed figures of the dielectric~response functions.}()}]{SM2023}%
  \BibitemOpen
  \bibfield  {author} {\bibinfo {author} {\bibfnamefont {See Supplemental~Material}\ \bibnamefont {at~[url will be inserted by publisher] for tables of the~irreducible representations}}\ and\ \bibinfo {author} {\bibnamefont {detailed figures of the dielectric~response functions.}},\ }\href@noop {} {\ }\BibitemShut {NoStop}%
\bibitem [{\citenamefont {Persson}\ and\ \citenamefont {Mirbt}(2006)}]{Persson2006}%
  \BibitemOpen
  \bibfield  {author} {\bibinfo {author} {\bibfnamefont {C.}~\bibnamefont {Persson}}\ and\ \bibinfo {author} {\bibfnamefont {S.}~\bibnamefont {Mirbt}},\ }\bibfield  {title} {\enquote {\bibinfo {title} {{Improved electronic structure and optical properties of sp-hybridized semiconductors using LDA+U$^{SIC}$}},}\ }\href@noop {} {\bibfield  {journal} {\bibinfo  {journal} {Br. J. Phys.}\ }\textbf {\bibinfo {volume} {36}},\ \bibinfo {pages} {286--290} (\bibinfo {year} {2006})}\BibitemShut {NoStop}%
\bibitem [{\citenamefont {Young}\ and\ \citenamefont {Frederikse}(1973)}]{Young1973}%
  \BibitemOpen
  \bibfield  {author} {\bibinfo {author} {\bibfnamefont {K.~F.}\ \bibnamefont {Young}}\ and\ \bibinfo {author} {\bibfnamefont {H.~P.~R.}\ \bibnamefont {Frederikse}},\ }\bibfield  {title} {\enquote {\bibinfo {title} {Compilation of the static dielectric constant of inorganic solids},}\ }\href@noop {} {\bibfield  {journal} {\bibinfo  {journal} {J. Phys. Chem. Ref. Data}\ }\textbf {\bibinfo {volume} {2}},\ \bibinfo {pages} {313--410} (\bibinfo {year} {1973})}\BibitemShut {NoStop}%
\bibitem [{\citenamefont {Bostr\"om}\ \emph {et~al.}()\citenamefont {Bostr\"om}, \citenamefont {Rizwan~Khan}, \citenamefont {Reddy~Gopidi}, \citenamefont {Brevik}, \citenamefont {Li}, \citenamefont {Persson},\ and\ \citenamefont {Malyi}}]{MathiasRizwanHarshanIverClasSashaPRB2023}%
  \BibitemOpen
  \bibfield  {author} {\bibinfo {author} {\bibfnamefont {M.}~\bibnamefont {Bostr\"om}}, \bibinfo {author} {\bibfnamefont {M.}~\bibnamefont {Rizwan~Khan}}, \bibinfo {author} {\bibfnamefont {H.}~\bibnamefont {Reddy~Gopidi}}, \bibinfo {author} {\bibfnamefont {I.}~\bibnamefont {Brevik}}, \bibinfo {author} {\bibfnamefont {Y.}~\bibnamefont {Li}}, \bibinfo {author} {\bibfnamefont {C.}~\bibnamefont {Persson}}, \ and\ \bibinfo {author} {\bibfnamefont {O.~I.}\ \bibnamefont {Malyi}},\ }\bibfield  {title} {\enquote {\bibinfo {title} {A knob to tune the {Casimir-Lifshitz} force with gapped metals},}\ }\href@noop {} {\bibinfo  {journal} {Phys. Rev. B (submitted)}\ }\BibitemShut {NoStop}%
\bibitem [{\citenamefont {Malyi}\ \emph {et~al.}(2016)\citenamefont {Malyi}, \citenamefont {Bostr{\"o}m}, \citenamefont {Kulish}, \citenamefont {Thiyam}, \citenamefont {Parsons},\ and\ \citenamefont {Persson}}]{Sasha2016}%
  \BibitemOpen
\bibfield  {journal} {  }\bibfield  {author} {\bibinfo {author} {\bibfnamefont {O.~I.}\ \bibnamefont {Malyi}}, \bibinfo {author} {\bibfnamefont {M.}~\bibnamefont {Bostr{\"o}m}}, \bibinfo {author} {\bibfnamefont {V.~V.}\ \bibnamefont {Kulish}}, \bibinfo {author} {\bibfnamefont {P.}~\bibnamefont {Thiyam}}, \bibinfo {author} {\bibfnamefont {D.~F.}\ \bibnamefont {Parsons}}, \ and\ \bibinfo {author} {\bibfnamefont {C.}~\bibnamefont {Persson}},\ }\bibfield  {title} {\enquote {\bibinfo {title} {{Volume dependence of the dielectric properties of amorphous SiO2}},}\ }\href@noop {} {\bibfield  {journal} {\bibinfo  {journal} {Phys. Chem. Chem. Phys.}\ }\textbf {\bibinfo {volume} {18}},\ \bibinfo {pages} {7483--7489} (\bibinfo {year} {2016})}\BibitemShut {NoStop}%
\bibitem [{\citenamefont {Fiedler}\ \emph {et~al.}(2020)\citenamefont {Fiedler}, \citenamefont {Bostr\"om}, \citenamefont {Persson}, \citenamefont {Brevik}, \citenamefont {Corkery}, \citenamefont {Buhmann},\ and\ \citenamefont {Parsons}}]{JohannesWater2019}%
  \BibitemOpen
  \bibfield  {author} {\bibinfo {author} {\bibfnamefont {J.}~\bibnamefont {Fiedler}}, \bibinfo {author} {\bibfnamefont {M.}~\bibnamefont {Bostr\"om}}, \bibinfo {author} {\bibfnamefont {C.}~\bibnamefont {Persson}}, \bibinfo {author} {\bibfnamefont {I.~H.}\ \bibnamefont {Brevik}}, \bibinfo {author} {\bibfnamefont {R.~W.}\ \bibnamefont {Corkery}}, \bibinfo {author} {\bibfnamefont {S.~Y.}\ \bibnamefont {Buhmann}}, \ and\ \bibinfo {author} {\bibfnamefont {D.~F.}\ \bibnamefont {Parsons}},\ }\bibfield  {title} {\enquote {\bibinfo {title} {Full-spectrum high resolution modeling of the dielectric function of water},}\ }\href@noop {} {\bibfield  {journal} {\bibinfo  {journal} {J.\ Phys.\ Chem.\ B}\ }\textbf {\bibinfo {volume} {124}},\ \bibinfo {pages} {3103--3113} (\bibinfo {year} {2020})}\BibitemShut {NoStop}%
\bibitem [{\citenamefont {Elbaum}\ and\ \citenamefont {Schick}(1991)}]{Elbaum}%
  \BibitemOpen
  \bibfield  {author} {\bibinfo {author} {\bibfnamefont {M.}~\bibnamefont {Elbaum}}\ and\ \bibinfo {author} {\bibfnamefont {M.}~\bibnamefont {Schick}},\ }\bibfield  {title} {\enquote {\bibinfo {title} {{Application of the theory of dispersion forces to the surface melting of ice}},}\ }\href@noop {} {\bibfield  {journal} {\bibinfo  {journal} {Phys. Rev. Lett.}\ }\textbf {\bibinfo {volume} {66}},\ \bibinfo {pages} {1713--1716} (\bibinfo {year} {1991})}\BibitemShut {NoStop}%
\bibitem [{\citenamefont {Li}\ \emph {et~al.}(2023)\citenamefont {Li}, \citenamefont {Corkery}, \citenamefont {Carretero-Palacios}, \citenamefont {Berland}, \citenamefont {Esteso}, \citenamefont {Fiedler}, \citenamefont {Milton}, \citenamefont {Brevik},\ and\ \citenamefont {Bostr\"om}}]{LiCorkeryCarreteroBerlandEstesoFiedlerMiltonBrevikBostrom2023}%
  \BibitemOpen
  \bibfield  {author} {\bibinfo {author} {\bibfnamefont {Y.}~\bibnamefont {Li}}, \bibinfo {author} {\bibfnamefont {R.~W.}\ \bibnamefont {Corkery}}, \bibinfo {author} {\bibfnamefont {S.}~\bibnamefont {Carretero-Palacios}}, \bibinfo {author} {\bibfnamefont {K.}~\bibnamefont {Berland}}, \bibinfo {author} {\bibfnamefont {V.}~\bibnamefont {Esteso}}, \bibinfo {author} {\bibfnamefont {J.}~\bibnamefont {Fiedler}}, \bibinfo {author} {\bibfnamefont {K.~A.}\ \bibnamefont {Milton}}, \bibinfo {author} {\bibfnamefont {I.}~\bibnamefont {Brevik}}, \ and\ \bibinfo {author} {\bibfnamefont {M.}~\bibnamefont {Bostr\"om}},\ }\bibfield  {title} {\enquote {\bibinfo {title} {Origin of anomalously stabilizing ice layers on methane gas hydrates near rock surface},}\ }\href@noop {} {\bibfield  {journal} {\bibinfo  {journal} {Phys. Chem. Chem. Phys.}\ }\textbf {\bibinfo {volume} {25}},\ \bibinfo {pages} {6636--6652} (\bibinfo {year} {2023})}\BibitemShut {NoStop}%
\bibitem [{\citenamefont {Esteso}\ \emph {et~al.}(2020)\citenamefont {Esteso}, \citenamefont {Carretero-Palacios}, \citenamefont {MacDowell}, \citenamefont {Fiedler}, \citenamefont {Parsons}, \citenamefont {Spallek}, \citenamefont {M\'iguez}, \citenamefont {Persson}, \citenamefont {Buhmann}, \citenamefont {Brevik},\ and\ \citenamefont {Bostr\"om}}]{Esteso4layerPCCP2020}%
  \BibitemOpen
  \bibfield  {author} {\bibinfo {author} {\bibfnamefont {V.}~\bibnamefont {Esteso}}, \bibinfo {author} {\bibfnamefont {S.}~\bibnamefont {Carretero-Palacios}}, \bibinfo {author} {\bibfnamefont {L.~G.}\ \bibnamefont {MacDowell}}, \bibinfo {author} {\bibfnamefont {J.}~\bibnamefont {Fiedler}}, \bibinfo {author} {\bibfnamefont {D.~F.}\ \bibnamefont {Parsons}}, \bibinfo {author} {\bibfnamefont {F.}~\bibnamefont {Spallek}}, \bibinfo {author} {\bibfnamefont {H.}~\bibnamefont {M\'iguez}}, \bibinfo {author} {\bibfnamefont {C.}~\bibnamefont {Persson}}, \bibinfo {author} {\bibfnamefont {S.~Y.}\ \bibnamefont {Buhmann}}, \bibinfo {author} {\bibfnamefont {I.}~\bibnamefont {Brevik}}, \ and\ \bibinfo {author} {\bibfnamefont {M.}~\bibnamefont {Bostr\"om}},\ }\bibfield  {title} {\enquote {\bibinfo {title} {Premelting of ice adsorbed on a rock surface},}\ }\href@noop {} {\bibfield  {journal} {\bibinfo  {journal} {Phys. Chem. Chem. Phys.}\ }\textbf {\bibinfo {volume} {22}},\ \bibinfo {pages} {11362--11373} (\bibinfo {year}
  {2020})}\BibitemShut {NoStop}%
\bibitem [{\citenamefont {Emelyanenko}\ \emph {et~al.}(2022)\citenamefont {Emelyanenko}, \citenamefont {Emelyanenko},\ and\ \citenamefont {Boinovich}}]{EmelyanenkoEmelyanenkoBoinovich2022}%
  \BibitemOpen
  \bibfield  {author} {\bibinfo {author} {\bibfnamefont {K.~A.}\ \bibnamefont {Emelyanenko}}, \bibinfo {author} {\bibfnamefont {A.~M.}\ \bibnamefont {Emelyanenko}}, \ and\ \bibinfo {author} {\bibfnamefont {L.~B.}\ \bibnamefont {Boinovich}},\ }\bibfield  {title} {\enquote {\bibinfo {title} {Review of the state of the art in studying adhesion phenomena at interfaces of solids with solid and liquid aqueous media},}\ }\href@noop {} {\bibfield  {journal} {\bibinfo  {journal} {Colloid J.}\ }\textbf {\bibinfo {volume} {84}},\ \bibinfo {pages} {265--286} (\bibinfo {year} {2022})}\BibitemShut {NoStop}%
\bibitem [{\citenamefont {Lebedew}(1894)}]{Lebedew1894}%
  \BibitemOpen
  \bibfield  {author} {\bibinfo {author} {\bibfnamefont {P.}~\bibnamefont {Lebedew}},\ }\bibfield  {title} {\enquote {\bibinfo {title} {{Ueber die mechanische Wirkung der Wellen auf ruhende Resonatoren. I. Electromagnetische Wellen}},}\ }\href@noop {} {\bibfield  {journal} {\bibinfo  {journal} {Annalen der Physik}\ }\textbf {\bibinfo {volume} {288}},\ \bibinfo {pages} {621--640} (\bibinfo {year} {1894})}\BibitemShut {NoStop}%
\bibitem [{\citenamefont {Derjaguin}\ \emph {et~al.}(2015)\citenamefont {Derjaguin}, \citenamefont {Abrikosova},\ and\ \citenamefont {Lifshitz}}]{Derjaguin_2015}%
  \BibitemOpen
  \bibfield  {author} {\bibinfo {author} {\bibfnamefont {B.~V.}\ \bibnamefont {Derjaguin}}, \bibinfo {author} {\bibfnamefont {I.~I.}\ \bibnamefont {Abrikosova}}, \ and\ \bibinfo {author} {\bibfnamefont {E.~M.}\ \bibnamefont {Lifshitz}},\ }\bibfield  {title} {\enquote {\bibinfo {title} {Molecular attraction of condensed bodies},}\ }\href@noop {} {\bibfield  {journal} {\bibinfo  {journal} {Physics-Uspekhi}\ }\textbf {\bibinfo {volume} {58}},\ \bibinfo {pages} {906} (\bibinfo {year} {2015})}\BibitemShut {NoStop}%
\bibitem [{\citenamefont {Parsegian}(2006)}]{Pars}%
  \BibitemOpen
  \bibfield  {author} {\bibinfo {author} {\bibfnamefont {V.~A.}\ \bibnamefont {Parsegian}},\ }\href@noop {} {\emph {\bibinfo {title} {Van der Waals forces: A handbook for biologists, chemists, engineers, and physicists}}}\ (\bibinfo  {publisher} {Cambridge University Press},\ \bibinfo {address} {New York},\ \bibinfo {year} {2006})\BibitemShut {NoStop}%
\bibitem [{\citenamefont {Ninham}\ and\ \citenamefont {Lo~Nostro}(2010)}]{Ninhb}%
  \BibitemOpen
  \bibfield  {author} {\bibinfo {author} {\bibfnamefont {B.~W.}\ \bibnamefont {Ninham}}\ and\ \bibinfo {author} {\bibfnamefont {P.}~\bibnamefont {Lo~Nostro}},\ }\href@noop {} {\emph {\bibinfo {title} {Molecular Forces and Self Assembly in Colloid, Nano Sciences and Biology}}}\ (\bibinfo  {publisher} {Cambridge University Press},\ \bibinfo {address} {Cambridge},\ \bibinfo {year} {2010})\BibitemShut {NoStop}%
\bibitem [{\citenamefont {Dou}\ \emph {et~al.}(2014)\citenamefont {Dou}, \citenamefont {Lou}, \citenamefont {Bostr\"om}, \citenamefont {Brevik},\ and\ \citenamefont {Persson}}]{PhysRevB.89.201407}%
  \BibitemOpen
  \bibfield  {author} {\bibinfo {author} {\bibfnamefont {M.}~\bibnamefont {Dou}}, \bibinfo {author} {\bibfnamefont {F.}~\bibnamefont {Lou}}, \bibinfo {author} {\bibfnamefont {M.}~\bibnamefont {Bostr\"om}}, \bibinfo {author} {\bibfnamefont {I.}~\bibnamefont {Brevik}}, \ and\ \bibinfo {author} {\bibfnamefont {C.}~\bibnamefont {Persson}},\ }\bibfield  {title} {\enquote {\bibinfo {title} {Casimir quantum levitation tuned by means of material properties and geometries},}\ }\href@noop {} {\bibfield  {journal} {\bibinfo  {journal} {Phys. Rev. B}\ }\textbf {\bibinfo {volume} {89}},\ \bibinfo {pages} {201407} (\bibinfo {year} {2014})}\BibitemShut {NoStop}%
\bibitem [{\citenamefont {Bostr\"om}\ and\ \citenamefont {Sernelius}(2000)}]{Bost2000}%
  \BibitemOpen
  \bibfield  {author} {\bibinfo {author} {\bibfnamefont {M.}~\bibnamefont {Bostr\"om}}\ and\ \bibinfo {author} {\bibfnamefont {Bo~E.}\ \bibnamefont {Sernelius}},\ }\bibfield  {title} {\enquote {\bibinfo {title} {{Thermal Effects on the Casimir Force in the 0.1-5\,$\mu$\,m Range}},}\ }\href@noop {} {\bibfield  {journal} {\bibinfo  {journal} {Phys. Rev. Lett.}\ }\textbf {\bibinfo {volume} {84}},\ \bibinfo {pages} {4757} (\bibinfo {year} {2000})}\BibitemShut {NoStop}%
\end{thebibliography}%
\end{document}



\title{Supplemental Material: Understanding ice and water film formation on soil particles by combining DFT and Casimir-Lifshitz forces}

\author{M.  Bostr{\"o}m}
\affiliation{Centre of Excellence ENSEMBLE3 Sp. z o. o., Wolczynska Str. 133, 01-919, Warsaw, Poland}

\author{S. Kuthe}
\affiliation{Department of Materials Science and Engineering, KTH Royal Institute of Technology, SE-100 44 Stockholm, Sweden}

 \author{S. Carretero-Palacios}
 \affiliation{Departamento de F\'isica de Materiales, Universidad Aut\'onoma de Madrid, 28049 Madrid, Spain}

 \author{V. Esteso}
\affiliation{Departamento de F\'isica de la Materia Condensada, ICMSE-CSIC, Universidad de Sevilla, Apdo. 1065, Sevilla, Spain}
\affiliation{European Laboratory for Non-Linear Spectroscopy (LENS), Via Nello Carrara 1, Sesto F.no 50019, Italy} 

\author{Y. Li}
  \affiliation{Department of Physics, Nanchang University, Nanchang 330031, China}
  \affiliation{Institute of Space Science and Technology, Nanchang University, Nanchang 330031, China}

\author{I. Brevik}
\affiliation{Department of Energy and Process Engineering, Norwegian University of Science and Technology, NO-7491 Trondheim, Norway}

\author{H. R. Gopidi}
\affiliation{Centre of Excellence ENSEMBLE3 Sp. z o. o., Wolczynska Str. 133, 01-919, Warsaw, Poland}
  
\author{O. I. Malyi}
\affiliation{Centre of Excellence ENSEMBLE3 Sp. z o. o., Wolczynska Str. 133, 01-919, Warsaw, Poland}


\author{B. Glaser}
\affiliation{Department of Materials Science and Engineering, KTH Royal Institute of Technology, SE-100 44 Stockholm, Sweden}

\author{C. Persson}
\affiliation{Department of Materials Science and Engineering, KTH Royal Institute of Technology, SE-100 44 Stockholm, Sweden}
\affiliation{Centre for Materials Science and Nanotechnology, Department of Physics, University of Oslo, P. O. Box 1048 Blindern, NO-0316 Oslo, Norway}

\maketitle


\par 
\begin{figure}[h!]
  \centering
  \includegraphics[width=0.7\columnwidth]{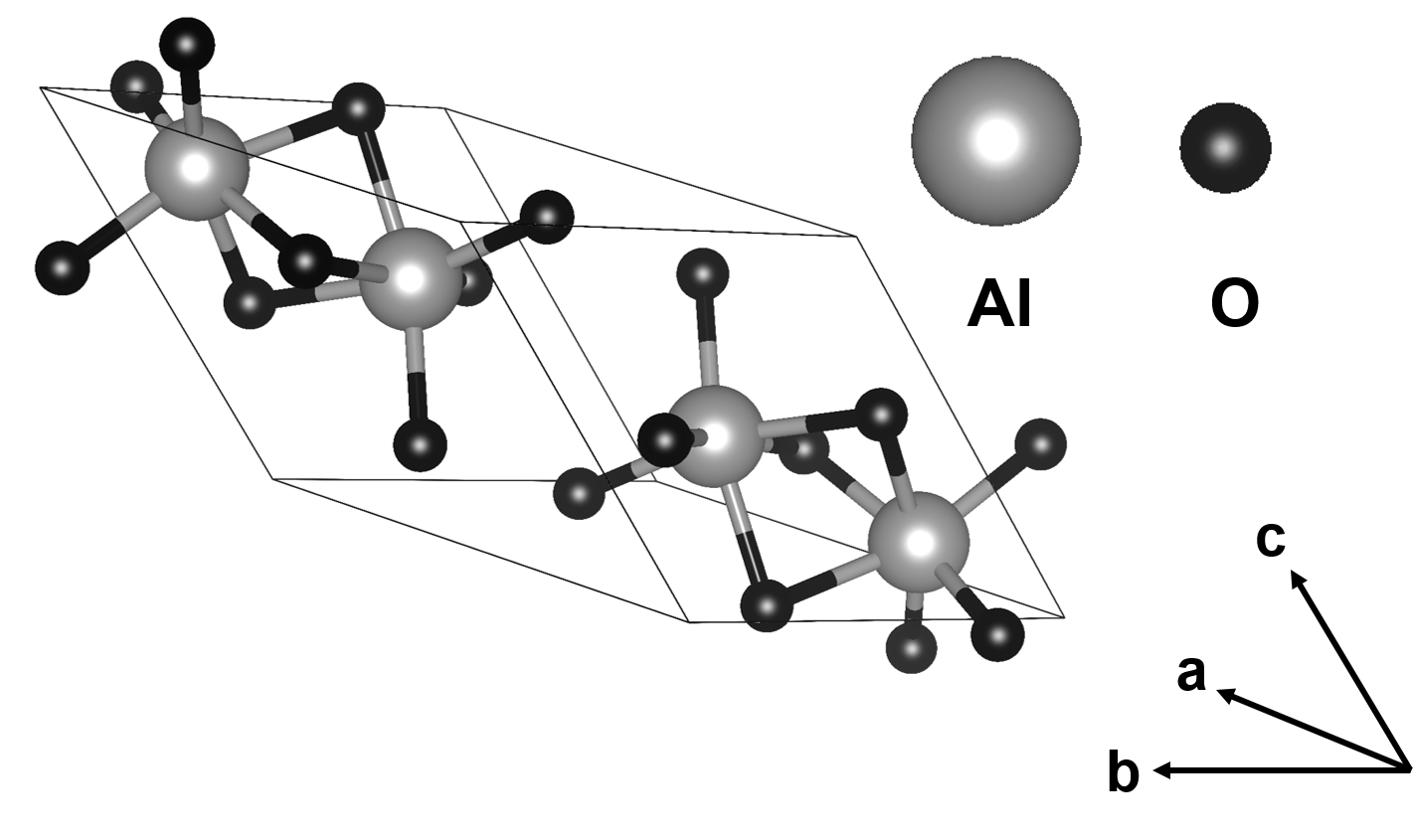}
  \caption{\label{Al2O3} 
  Structure of Al$_2$O$_3$}
  \label{fig:dft_Al2O3}
\end{figure}

\begin{figure}[h!]
  \centering
  \includegraphics[width=0.7\columnwidth]{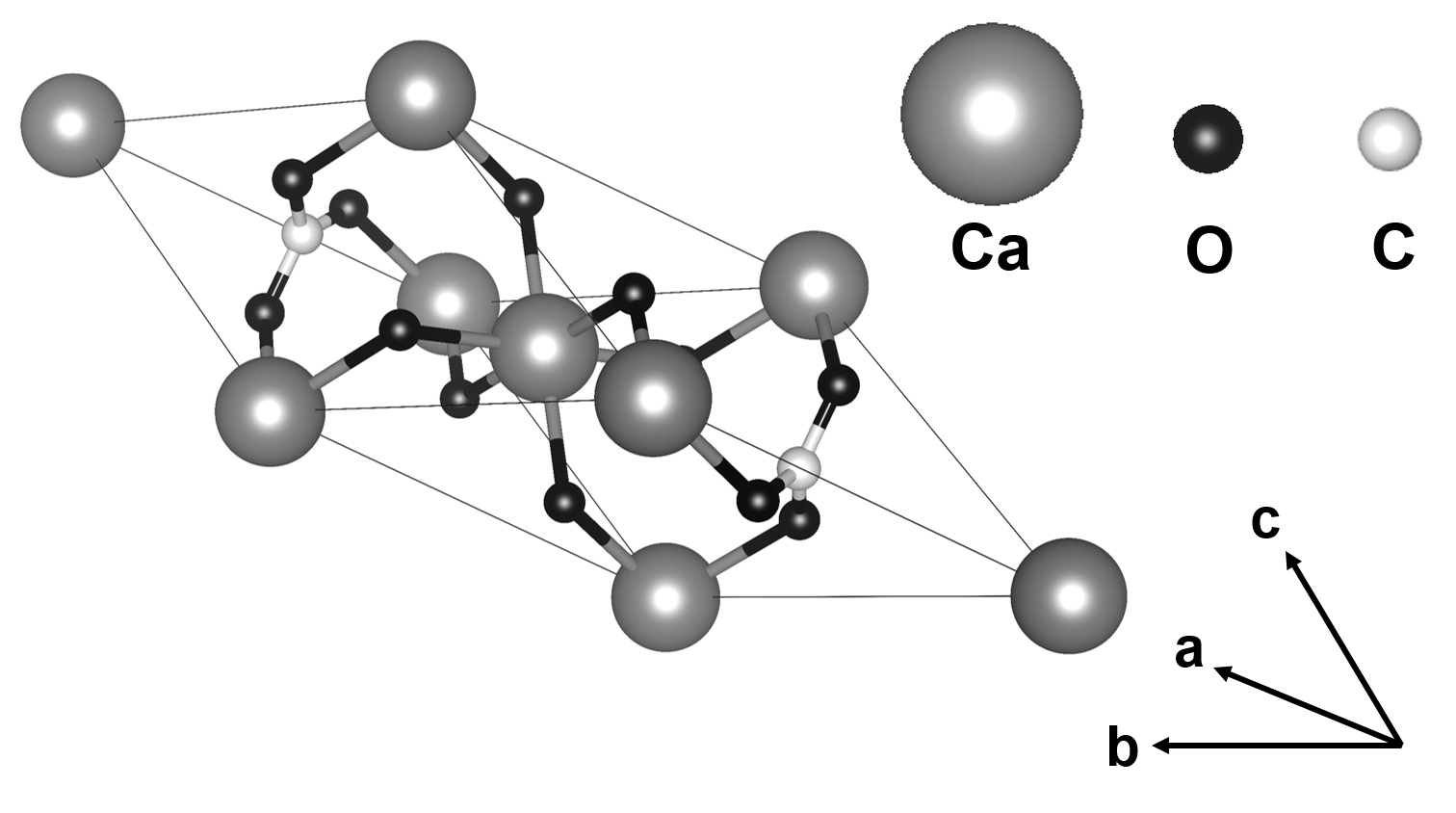}
  \caption{\label{CaCO3} 
  Structure of CaCO$_3$}
  \label{fig:dft_CaCO3}
\end{figure}
\par 

\newpage

\begin{table}[h!]
\caption{Atom positions of Al$_2$O$_3$ for the rhombohedral lattice and with the PBEsol potential.}
\centering
\begin{tabular}{|c|c|c|c|c|}
\hline
No. & Atom & $x$ & $y$ & $z$ \\
\hline
1 & Al & 0.14794 & 0.14794 & 0.14794  \\
2 & Al  & 0.85206 & 0.85206 & 0.85206 \\
3 & Al  & 0.35206 & 0.35206 & 0.35206  \\
4 & Al  & 0.64794 & 0.64794 & 0.64794  \\
5 & O  & 0.25000 & 0.55585 & 0.94415  \\
6 & O  & 0.75000 & 0.44415 & 0.05585 \\
7 & O  & 0.94415 & 0.25000 & 0.55585 \\
8 & O  & 0.05585 & 0.75000 & 0.44415 \\
9 & O & 0.55585 & 0.94415 & 0.25000  \\
10 & O & 0.44415 & 0.05585 & 0.75000 \\
\hline
\end{tabular}
\label{tab:structure_parameters1}
\end{table}

\begin{table}[h!]
\caption{Corresponding data for CaCO$_3$ as in previous table.}
\centering
\begin{tabular}{|c|c|c|c|c|}
\hline
No. & Atom & $x$ & $y$ & $z$  \\
\hline
1 & Ca  & 0.50000 & 0.50000 & 0.50000  \\
2 & Ca  & 0.00000 & 0.00000 & 0.00000 \\
3 & C & 0.75000 & 0.75000 & 0.75000  \\
4 & C   & 0.25000 & 0.25000 & 0.25000  \\
5 & O    & 0.75000 & 0.49132 & 0.00868  \\
6 & O  & 0.00868 & 0.75000 & 0.49132  \\
7 & O  & 0.49132 & 0.00868 & 0.75000  \\
8 & O    & 0.25000 & 0.50868 & 0.99132  \\
9 & O    & 0.99132 & 0.25000 & 0.50868  \\
10 & O & 0.50868 & 0.99132 & 0.25000  \\
\hline
\end{tabular}
\label{tab:structure_parameters2}
\end{table}

\newpage
\par 

\begin{table}[ht!]
\caption{The  electronic energies of Al$_2$O$_3$ and their single-group irreducible representations at at three symmetry points around the band edge. 
Energies are in units of eV, and the $\Gamma$-point valence state maximum has been set to zero in energy. The conduction band states have been shifted to agree with the $\Gamma$-point energy gap from the hybrid functional calculation with 30\% Hartree-Fock exchange.}
 \renewcommand{\arraystretch}{1.2}
\begin{tabular}{ lr r cr r cr } 
 \hline  \hline 
  $\Gamma$  &&\hspace{0.5cm}\ &  T &&\hspace{0.5cm}\ & L & \\
 \hline 
$\Gamma_{2}^-$   &$13.28$ &&  $T_3$ & $13.60$ && $L_1$ & $12.17$ \\
$\Gamma_{1}^-$   &$11.60$ &&            &             &&           &             \\
$\Gamma_{1}^+$  &$ 8.71$ &&  $T_3$ &  $11.17$ && $L_1$ & $11.96$ \\
\\
$\Gamma_{2}^-$   &$ 0.00$ &&  $T_1+T_2$ & $-0.38$    && $L_1$ & $-0.27$ \\
$\Gamma_{3}^-$   &$-0.04$ &&                    &               && $L_1$ & $-0.91$ \\
$\Gamma_{3}^+$  &$-0.50$ &&  $T_1+T_2$ &  $-2.01$  && $L_1$ & $-1.31$ \\
$\Gamma_{3}^+$  &$-1.29$ &&                    &               && $L_1$ &  $-1.55$\\
$\Gamma_{3}^-$   &$-1.41$ &&  $T_3$          & $-2.27$  && $L_1$ &  $-2.21$\\ 
$\Gamma_{2}^-$   &$-2.96$ &&                     &              &&          &  \\
\hline  \hline  
\end{tabular} 
\label{tab:irrepal2o3}
\end{table}

\begin{table}[ht!]
\caption{Corresponding data for CaCO$_3$ as in previous table.}
 \renewcommand{\arraystretch}{1.2}
\begin{tabular}{ lr r cr r cr }
 \hline  \hline 
  $\Gamma$  &&\hspace{0.5cm}\ &  T &&\hspace{0.5cm}\ & L & \\ 
 \hline 
$\Gamma_{2}^+$  &$ 8.08$ &&  $T_1+T_2$ &  $9.64$ && $L_1$ & $9.52$ \\
$\Gamma_{2}^-$   &$ 8.01$ &&                    &             &&           &             \\
$\Gamma_{1}^+$  &$ 7.97$ &&  $T_3$         &  $7.63$ && $L_1$ & 8.10 \\
\\
$\Gamma_{2}^-$   &$ 0.00$ &&  $T_3$         & $-0.27$    && $L_1$ & $0.10$ \\
$\Gamma_{2}^+$  &$-0.48$ &&                    &               &&           &  \\
$\Gamma_{3}^-$   &$-0.62$ &&  $T_1+T_2$  &  $-0.99$  && $L_1$ & $-1.15$ \\
$\Gamma_{3}^-$   &$-1.26$ &&                    &               && $L_1$ &  $-1.42$\\
$\Gamma_{3}^+$  &$-1.51$ &&  $T_1+T_2$  & $-1.71$  && $L_1$ &  $-1.72$\\ 
$\Gamma_{3}^+$  &$-1.95$ &&                     &              && $L_1$ &  $-2.13$\\
 \hline  \hline  
\end{tabular} 
\label{tab:irrepcaco3}
\end{table}

\newpage

\begin{figure}[h!]
  \centering
  \includegraphics[width=0.63\columnwidth]{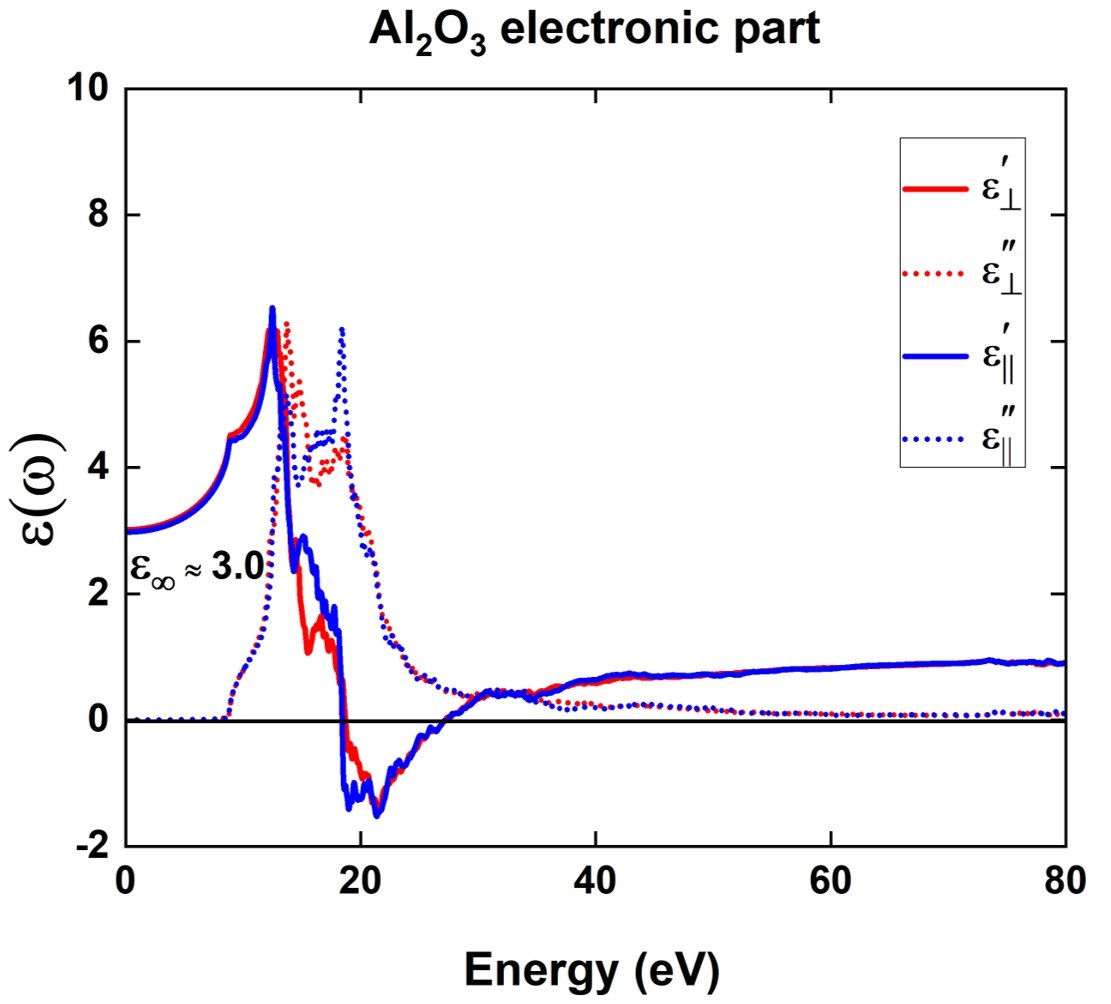}
  \caption{\label{CaCO3 structure} The electronic contributions for Al$_2$O$_3$.}
  \label{fig:dft_elec_Al2O3}
\end{figure}

\begin{figure}[h!]
  \centering
  \includegraphics[width=0.63\columnwidth]{Figures/CaCO3_Electronic part.jpg}
  \caption{\label{CaCO3 structure} The electronic contributions for CaCO$_3$.}
  \label{fig:dft_elec_CaCO3}
\end{figure}

\newpage
\begin{figure}[h!]
  \centering
  \includegraphics[width=0.7\columnwidth]{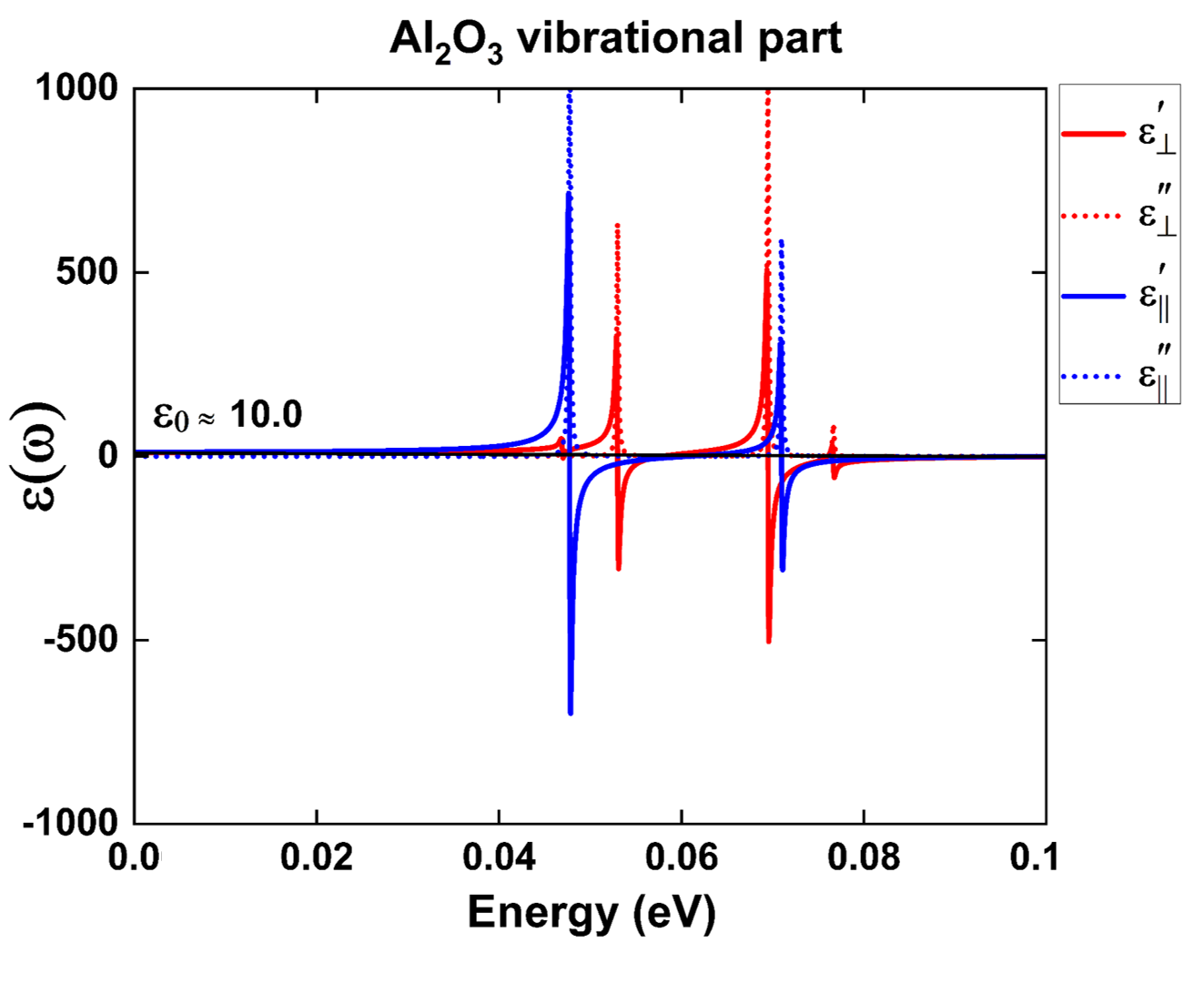}
  \caption{\label{CaCO3 structure}  The vibrational contributions for Al$_2$O$_3$.}
  \label{fig:dft_vib_Al2O3}
\end{figure}

\begin{figure}[h!]
  \centering
  \includegraphics[width=0.7\columnwidth]{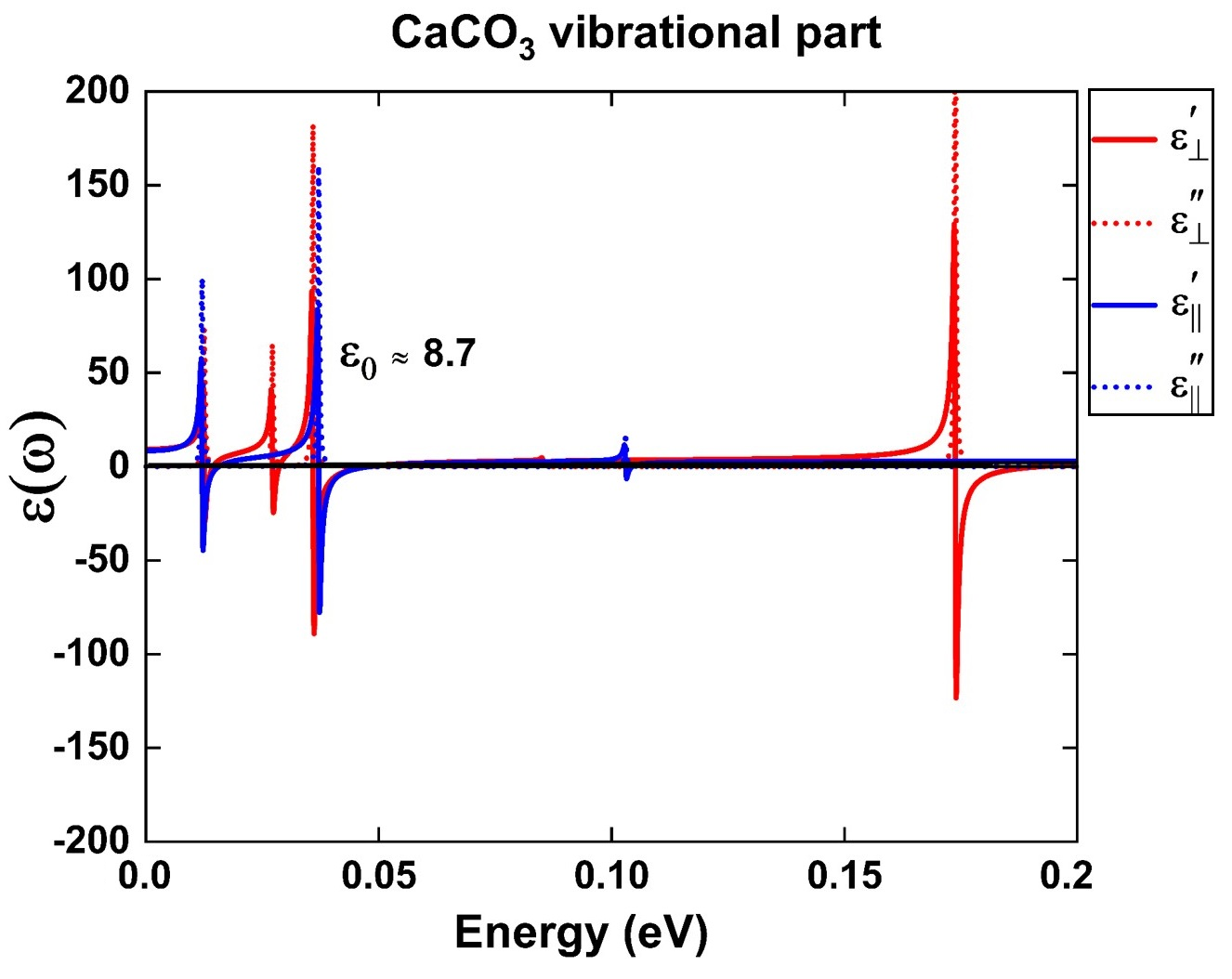}
  \caption{\label{CaCO3 structure} The vibrational contributions for CaCO$_3$.}
  \label{fig:dft_vib_CaCO3}
\end{figure}